\documentclass[twocolumn]{aastex63}

\usepackage{gensymb}

\newcommand{\e}{$e^{-}$}

\newcommand{\nw}{nW m$^{-2}$ sr$^{-1}$}

\shorttitle{Reconstruction of Severely Undersampled PSFs}
\shortauthors{Symons et al.}
\accepted{for publication in \apjs}

\begin{document}

\title{Superresolution Reconstruction of Severely Undersampled Point-spread Functions Using Point-source Stacking and Deconvolution}

\correspondingauthor{Teresa Symons}
\email{tas4514@rit.edu}

\author[0000-0002-9554-1082]{Teresa Symons}
\affiliation{Center for Detectors, Rochester Institute of Technology, 1 Lomb Memorial Drive, Rochester, NY 14623, USA}

\author[0000-0001-8253-1451]{Michael Zemcov}
\affiliation{Center for Detectors, Rochester Institute of Technology, 1 Lomb Memorial Drive, Rochester, NY 14623, USA}
\affiliation{Jet Propulsion Laboratory, California Institute of Technology, Pasadena, CA 91109, USA}

\author{James Bock}
\affiliation{Jet Propulsion Laboratory, California Institute of Technology, Pasadena, CA 91109, USA}
\affiliation{California Institute of Technology, 1200 E. California Boulevard, Pasadena, CA 91125, USA}

\author[0000-0002-5437-0504]{Yun-Ting Cheng}
\affiliation{California Institute of Technology, 1200 E. California Boulevard, Pasadena, CA 91125, USA}

\author[0000-0002-4650-8518]{Brendan Crill}
\affiliation{Jet Propulsion Laboratory, California Institute of Technology, Pasadena, CA 91109, USA}

\author{Christopher Hirata}
\affiliation{Center for Cosmology and AstroParticle Physics, The Ohio State University, 191 West Woodruff Avenue, Columbus, OH 43210, USA}

\author{Stephanie Venuto}
\affiliation{Center for Detectors, Rochester Institute of Technology, 1 Lomb Memorial Drive, Rochester, NY 14623, USA}


\begin{abstract}
Point-spread function (PSF) estimation in spatially undersampled images is challenging because large pixels average fine-scale spatial information. This is problematic when fine-resolution details are necessary, as in optimal photometry where knowledge of the illumination pattern beyond the native spatial resolution of the image may be required. Here, we introduce a method of PSF reconstruction where point sources are artificially sampled beyond the native resolution of an image and combined together via stacking to return a finely sampled estimate of the PSF. This estimate is then deconvolved from the pixel-gridding function to return a superresolution kernel that can be used for optimally weighted photometry. We benchmark against the $<$1\% photometric error requirement of the upcoming SPHEREx mission to assess performance in a concrete example. We find that standard methods like Richardson--Lucy deconvolution are not sufficient to achieve this stringent requirement. We investigate a more advanced method with significant heritage in image analysis called iterative back-projection (IBP) and demonstrate it using idealized Gaussian cases and simulated SPHEREx images. In testing this method on real images recorded by the LORRI instrument on New Horizons, we are able to identify systematic pointing drift. Our IBP-derived PSF kernels allow photometric accuracy significantly better than the requirement in individual SPHEREx exposures. This PSF reconstruction method is broadly applicable to a variety of problems and combines computationally simple techniques in a way that is robust to complicating factors such as severe undersampling, spatially complex PSFs, noise, crowded fields, or limited source numbers.
\end{abstract}


\section{Introduction}
\label{S:intro}

\setcounter{footnote}{4}

Whether explicitly or implicitly, astrophysicists require knowledge of how the brightness in an image relates to true brightness on the sky to 
interpret the shapes and intensities of observed sources of
emission. The instrument response function of a telescope is the transfer function that maps between quantities measured
 at the focal plane to the physical intensity on the sky. For natively imaging array detectors, or images constructed
from time-domain scans of the sky, this information is often called the
point-spread function (PSF), which is defined to be the response of a focused
imaging system to a point source (see \citealt{Gai2007} for a review). PSF estimation is a foundational problem in astronomical image analysis and interpretation.

Measuring the PSF of an instrument can be challenging.
For some instruments, it is possible to measure the PSF using
collimated images of unresolved sources in the laboratory or on the sky. It is often possible to use geometric or physical optics to model the system and propagate an image of an
unresolved source to the detector. More recently, forward-modeling approaches incorporating measurements of the PSF have been used to model complex electrical effects in detectors \citep[e.g.,][]{Donlon2018}. In general,
astrophysicists have a preference for using images of unresolved\footnote{For our purposes, ``unresolved'' is the condition
$\theta_{\rm source} \ll \theta_{\rm PSF}$, where $\theta_{\rm source}$ is the width of the source and $\theta_{\rm PSF}$ is the width of the PSF.}
objects to assess the PSF in an image. This method has the advantage
of measuring the as-built optical system, which not only captures the design performance of an optical system, but also any effects from scattered light, mechanical structures, or
other nonidealities in the telescope or detector.

\begin{figure*}[htbp!]
\centering
\includegraphics{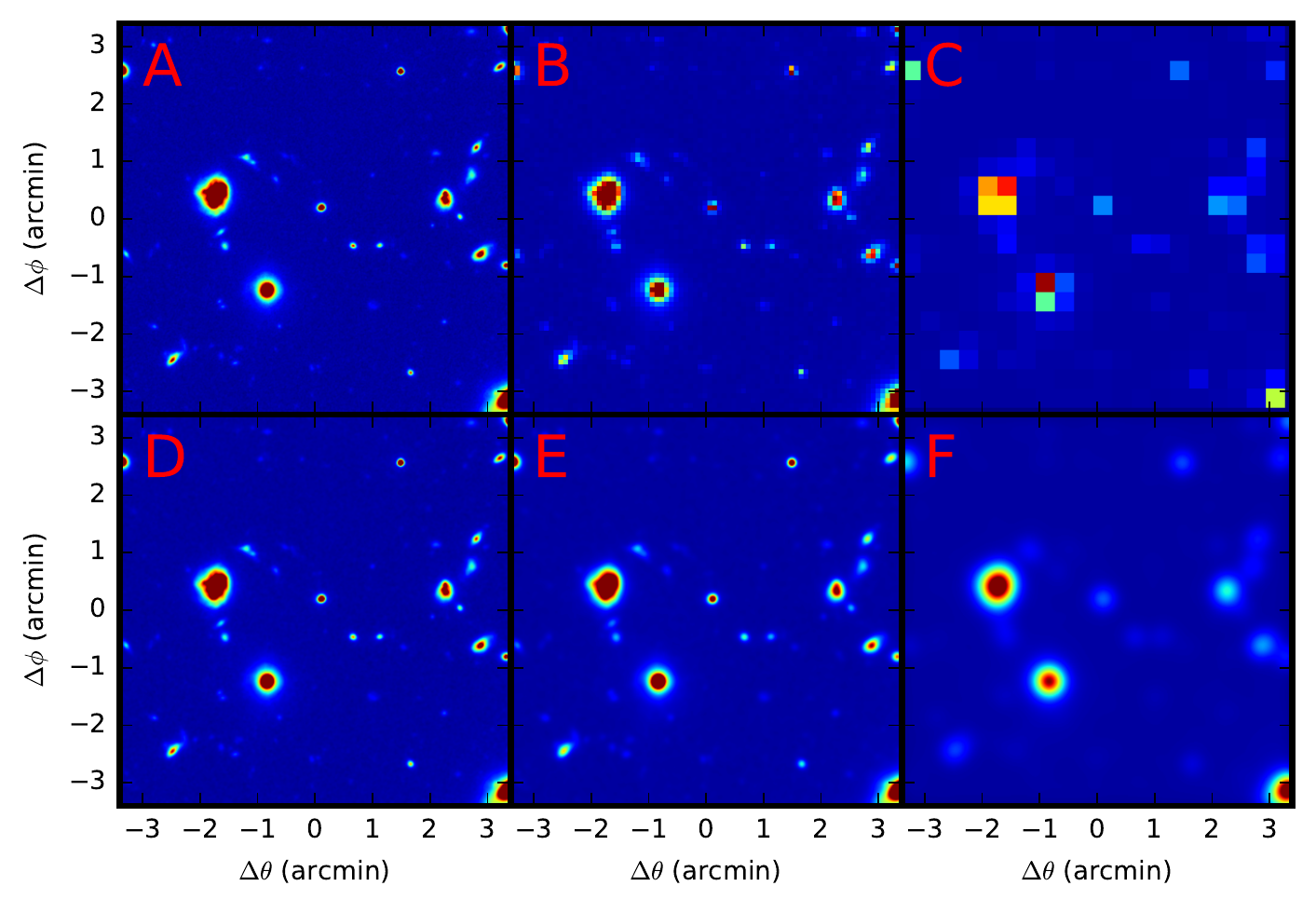}
\caption{Examples of the relationship between pixelization and PSF.
 The top row shows the effect of changes in the pixel gridding of the
 input image, shown in panel A, while the bottom row shows the
 effect of changing the width of the PSF. In panel A, the pixelization is matched
 to the optical PSF, so that the FWHM $\sim$1 pixel. In this case,
 the spatial resolution of the telescope dominates the spatial
 resolution of the image. In panel B, we show the case where
 $\theta_{\rm pix} $/FWHM $\sim$5 and the image spatial resolution is
 dominated by the pixel grid. Panel C shows the
 $\theta_{\rm pix} $/FWHM $\sim$20 case where the image spatial
 resolution is heavily gridding-dominated. The method described here
 takes advantage of the fact that the PSF is sampled in many
 different ways with respect to the pixel grid to allow
 reconstruction of the subpixel PSF shape. In the bottom row, we
 show examples of FWHM$_{\rm PSF} / \theta_{\rm pix} = \{2, 5, 20\}$
 in panels D, E, and F, respectively. In these cases, the subpixel
 PSF can be easily measured from point-like sources, and the method
 described here offers no improvement. \label{fig:resolution} }
\end{figure*}

The intrinsic PSF of the optical
system and the pixelization of the displayed image are not immutable quantities, but rather independent
parameters fixed at either the instrument design, or for the case of
time-domain maps, during the map-making process. This is illustrated in Figure
\ref{fig:resolution}, which shows the effects of changing the relative sizes of the pixelization and PSF. In practice, the choice of image pixelization depends on the sources
under investigation and the desired properties of the final image.
When possible, the pixel size is chosen to be slightly smaller than the intrinsic PSF,
but not to ``superresolve'' it. This is because of both the noise penalties
associated with spreading a fixed flux into an increased number of pixels, and the diminishing fidelity of superresolved astronomical images. A resolution of two pixels per FWHM is often quoted as representative of Nyquist sampling, with a smaller number of pixels per FWHM indicating undersampling. If the optical PSF is significantly narrower than the width of a pixel, then the PSF is severely undersampled and the precise location of the source within the pixel may be difficult to recover \citep{dig_im_recon, psf_sampling}. It can be shown that the optimum signal-to-noise ratio (S/N) on unresolved sources is achieved by either concentrating as much of the optical PSF into a single detector element as possible, or by performing optimal photometry that uses knowledge of the PSF to weight pixels according to their flux contributions \citep[e.g.,][]{Horne1986, Naylor1998}. The need for accurate photometry of sources is a strong motivation to improve our knowledge of the spatial properties of the PSF, sometimes beyond that offered by the native image pixelization.

Though the optical PSF may not be resolved by the image pixelization,
information about its shape must be retained in any ensemble of images of
unresolved sources that are placed at random locations with respect to
the pixel grid. In this situation, the images of the PSF are sampled by
the pixel grid slightly differently for each member of the set, so with a sufficient number of draws it should be possible to recombine the information to reconstruct the underlying ``unpixelized''
optical PSF. This concept has been successfully used to reconstruct superresolved PSFs in the past using a method commonly referred to as ``stacking'' \citep{Dole2006,Bethermin2012,Zemcov2014}. In short, the stacking method is a measurement of the covariance between a catalog of sources and their corresponding images in a map. Superresolution\footnote{We define superresolution to mean any resolution higher than an image's native resolution.} methods such as stacking can offer a way to recover Nyquist sampling of the PSF \citep{decon_review}. The stacking method \citep{shiftandadd} was
first used in this context to perform statistical studies on low-resolution, confused
images of the cosmic infrared background (e.g.,
\citealt{Dole2006,Marsden2009}, and others), but has since been extended to a wide range of wavelengths and scientific topics \citep[see, e.g.,][]{Viero2013}. 

In this paper, we study PSF reconstruction and modeling using the stacking technique. We demonstrate that,
in the limit where the input catalog is composed of point sources, the image resulting from a stacking process is that of the intrinsic
PSF convolved with the pixelization grid (the effective PSF), and that the pixel grid can be reliably deconvolved to retrieve the instrument's intrinsic optical PSF. 
As a case study, we examine this PSF reconstruction algorithm
as it may apply to the Spectro-Photometer for the History of the universe, Epoch of
Reionization, and ices Explorer (SPHEREx), an upcoming spectroscopic
survey mission designed to image every $6.2 \times 6.2$
arcsec$^{2}$ pixel over the entire sky for
0.75 $<\lambda<$ 4.8 $\mu$m. Accurate photometry with methodological uncertainties $<$1\% is required to help attain SPHEREx's ambitious scientific goals, but the PSF is intentionally designed to be underresolved by the pixel grid with an optical FWHM of $2.^{\prime \prime}01$ at 0.75 $\mu$m, which is consistent with \cite{Korngut2018} when manufacturing tolerance and aberrations are included. This gives a sampling rate of $\sim$0.3 pixels per FWHM, well below the Nyquist rate. It is therefore necessary to construct an algorithm that provides accurate PSF kernels for the precision optimal photometry. Because of SPHEREx's severe undersampling, PSF reconstruction techniques used in other cases where accurate spatial knowledge of the PSF is equally important, such as for the study of weak lensing, cannot be used here \citep{psfex, imcom, euclid_psf}. These methods typically require applying subpixel shifts and thus some form of interpolation, but in our method, all shifts are in the form of integer values due to the large upsampling factor. The proposed approach also offers the benefit of only requiring a single exposure instead of combining multiple exposures to reconstruct the PSF and is less computationally expensive \citep{imcom_test}. The SPHEREx case illustrates the utility of our method in a situation where the intrinsic PSF is unresolved in individual images of point
sources but can be reconstructed from a large ensemble of measurements contained within a single exposure. 
In Section \ref{S:theory}, we introduce the PSF
stacking concept in detail, various methods of deconvolution necessary for reconstructing the PSF in complex cases, as well as the method by which the PSF reconstruction can be used as a weight function in calculating photometry. In Section \ref{S:results}, we apply these methods to example cases including simulated SPHEREx PSFs as well as real data from the Long Range Reconnaissance Imager (LORRI) on New Horizons \citep{Cheng2008}. In Section \ref{S:discussion}, we put our work in the context of generalized PSF reconstruction techniques. Table \ref{table:symtab} provides a glossary of mathematical variables used throughout the text.

\vspace{10pt}
\section{Methods}
\label{S:theory}

\begin{deluxetable*}{cl}[htbp!]
\tablecaption{Important Variables Used in the Text\label{table:symtab}}
\tablehead{\colhead{\bf{Variable}} & \colhead{\bf{Description}}}
\centering
\startdata
 \multicolumn{2}{c}{\textit{Image quantities, native pixelization}} \\ \hline
 $M$ & sky image \\
 $\Delta$ & noise in $M$ \\
 $P_{\rm true}$ & underlying optical PSF of system\\
 $P_{\rm grid}$ & pixel grid function\\
 $P$ & the PSF of sources in $M$, convolution of $P_{\rm true}$ and $P_{\rm grid}$ \\ \hline
 \multicolumn{2}{c}{\textit{Image quantities, super pixelization}} \\ \hline
 $r$ & pixel grid scaling factor \\
 $\tilde{M}$ & sky image $M$ scaled up by $r$ \\
 $\tilde{P}$ & PSF of sources in $M$ scaled up by $r$ \\ \hline
 \multicolumn{2}{c}{\textit{Input source catalog quantities}} \\ \hline
 $\alpha$ & list of sources in image $M$ \\
 $F$ & known flux of sources in $M$ \\
 $(x,y)$ & known positions of sources in $M$ \\
 $N$ & number of sources in $M$ \\ \hline
\multicolumn{2}{c}{\textit{PSF reconstruction quantities}} \\ \hline
 $T$ & thumbnail cutout of a source centered on its known position \\
 $P_{\rm S}$ & stacked PSF \\
 $P_{\rm RLD}$ & deconvolution of $P_{\rm S}$ via RLD \\
 $P_{\rm IBP}$ & deconvolution of $P_{\rm S}$ via IBP, best estimate for $P_{\rm true}$ \\
 $I_{\rm RLD}$ & number of RLD iterations \\
 $I_{\rm IBP}$ & number of IBP iterations \\ \hline
 \multicolumn{2}{c}{\textit{Optimal photometry quantities}} \\ \hline
 $P_{\rm P}$ & recentered, rescaled version of $P_{\rm IBP}$ used in optimal photometry \\
 $W_{\rm P}$ & weight function used for optimal photometry \\
 $F_{\rm P}$ & flux of a source calculated from optimal photometry \\
 $\langle \delta F \rangle$ & average deviation of flux from its known value \\
 $\sigma_{\delta F}$ & PSF reconstruction figure of merit \\
\enddata
\end{deluxetable*}

\subsection{Stacking}
\label{S:stack}

\begin{figure*}[htbp!]
\centering
\includegraphics{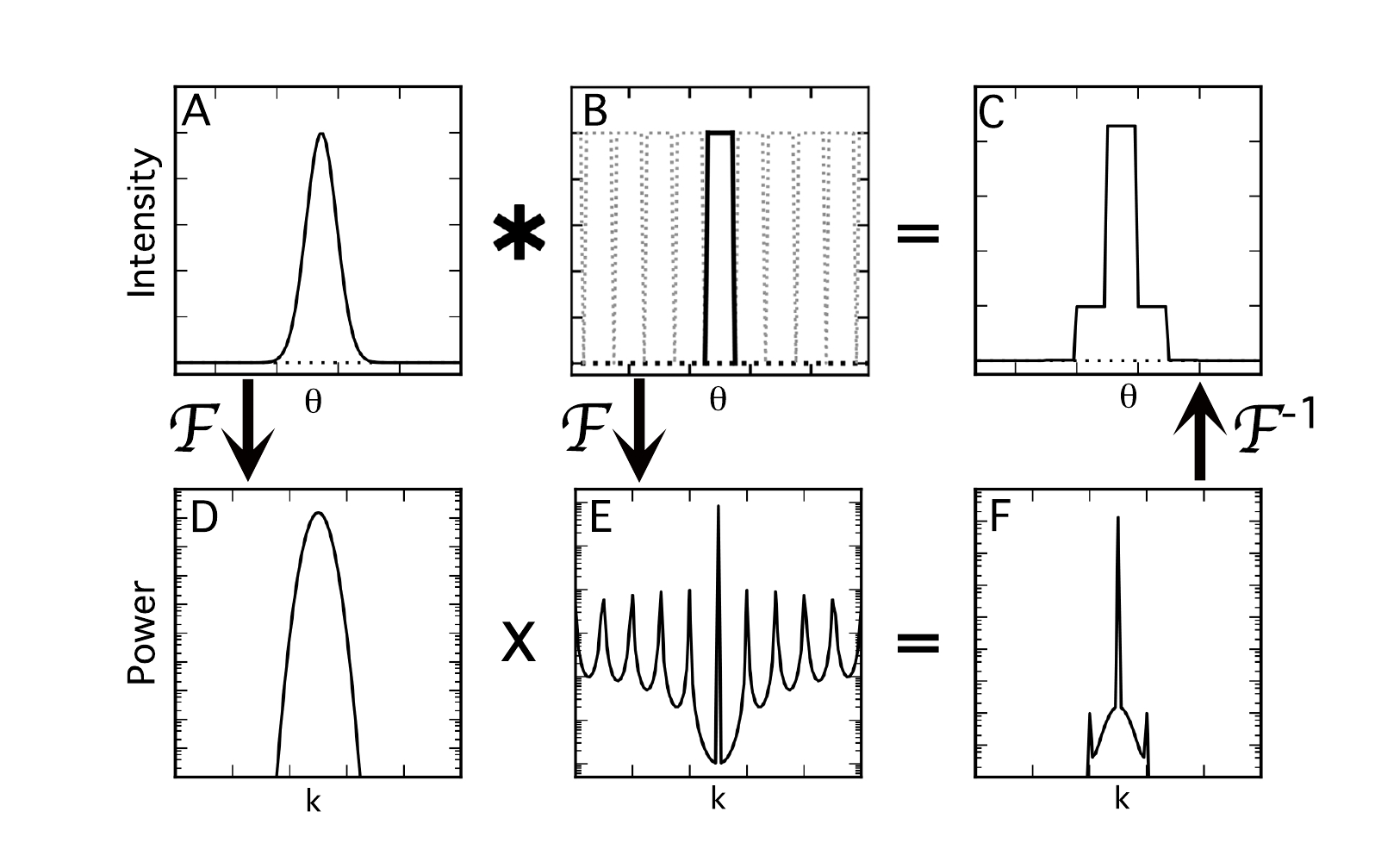}
\caption{Relationship of the PSF and pixel-gridding function in both
 real space and Fourier space. Panel A shows a hypothetical PSF,
 which has a similar size to the pixel-gridding function (dotted line) shown in panel
 B. The spatial resolution of the image that results is computed as
 the convolution of the PSF and a single pixel (solid line in panel B), as shown in
 panel C. Flux from a point source is spread across several pixels
 according to the brightness in the PSF as sampled by the pixel
 gridding function. As an alternative visualization, panel D shows
 the power spectrum (absolute square of the Fourier transform) of
 the PSF, and panel E shows the power spectrum of the pixel-gridding
 function. By the convolution theorem, the inverse Fourier
 transform of their product, shown as a power spectrum in panel F,
 is also the final spatial quality of the image of a point source.
 These two visualizations are useful for understanding interactions
 between the PSF and pixel-gridding function. \label{fig:convolution} }
\end{figure*}

A simple way to understand the relationship between the PSF and the
pixel grid function is through the convolution theorem. 
The act of signal detection with a pixelized system is equivalent to a convolution of the true underlying optical PSF, $P_{\rm true}$, with the pixel response $P_{\rm grid}$: 
\begin{equation}
P = P_{\rm true} \ast P_{\rm grid} = \mathcal{F}^{-1}[ \mathcal{F}[P_{\rm true}] \cdot \mathcal{F}[P_{\rm grid}]],
\end{equation}
where $\mathcal{F}$ represents the Fourier transform and $P$ is the PSF that can be observed in the image itself \citep{decon_review,dig_im_recon}, sometimes called the effective PSF \citep{Lauer1999}. 
Figure \ref{fig:convolution} illustrates the relationship between these quantities.

In our formulation, the quantity $P_{\rm true}$ is the combination of any effects that contribute to the overall PSF measured in an image. These include the underlying optical PSF, how the optical PSF is played around pixels due to pointing jitter, and any electrical effects like interpixel capacitance or the ``brighter-fatter'' effect \citep{Hirata2020}. For a linear detector, the quantity $P_{\rm true}$ incorporates all of these effects as a convolution over the history of the exposure. It is commonly noted that the observer only ever has access to $P$, so it is expedient to perform analysis with this quantity. But in the presence of nonlinear effects that are known to exist in modern hybridized HgCdTe NIR detectors \citep{Plazas2018, Hirata2020}, it may in fact be necessary to deconvolve $P_{\rm grid}$ or other effects from $P$. There are potentially many reasons and ways to do this; here, we investigate one that uses the fact that the covariance between a well-understood catalog of stars and sources in the image allows for superresolution reconstruction, which when combined with deconvolution methods can return the underlying optical PSF. This is particularly motivated by the fact that, in the presence of these complicating factors, $P_{\rm true}$ is the kernel required to perform optimal photometry. 
 
The stacking method takes advantage of the fact that an ``ideal''
gridded image of the astronomical sky can be represented as
\begin{equation}
\label{eq:stacking}
M_{j} = \Delta_{j} + \sum_{i}^{N_{j}} F_{i},
\end{equation}
where $M_{j}$ is the brightness of the image $M$ in pixel $j$, $\Delta_{j}$ is the noise in that pixel, $N_{j}$ is the number of sources
falling into pixel $j$, and $F_{i}$ is the flux of source $i$ in a list $\alpha$ of $N$ total sources of emission
\citep{Marsden2009,Viero2013}. To make progress with PSF estimation,
we note that, by the argument above, instruments with non-point PSFs
contribute flux into more than one pixel. We define
$P_{j} (x_{i},y_{i})$ to be the beam-convolved and mean-subtracted shape of the PSF centered at some source position
$(x_{i},y_{i})$, and then write Equation \ref{eq:stacking}
as
\begin{equation}
\label{eq:betterstacking}
M_{j} = \Delta_{j} + \sum_{i}^{N_{j}} F_{i} P_{j}(x_{i},y_{i}).
\end{equation}
This expression accounts for the fact that all sources can contribute
to the intensity of pixel $j$, as the PSF spreads
flux to neighboring pixels.

Our goal is to estimate the shape of the PSF $P$ from the measured sky
$M$. Because $P$ has the same amplitude for each source, we can
invert Equation \ref{eq:betterstacking} and solve
\begin{equation}
\label{eq:psfstack}
P = \sum_{i}^{N_{j}} \frac{M(x_{i},y_{i})}{F_{i}},
\end{equation}
where $M(x_{i},y_{i})$ is the image centered on the source
position $(x_{i},y_{i})$, and we assume that the noise obeys
$\langle \Delta \rangle = 0$ over the sum. Furthermore, as with simple
stacking, we require the source positions to be uncorrelated so that
contributions from sources in the image but not being stacked on do
not add coherently \citep{Marsden2009}.

In this formalism, the superresolution recovery arises due to
the nature of pixels, which have the property of averaging photons.
We can always create
larger pixels which contain more photons, but spread over a larger
area, in such a way that the measured surface brightness is
conserved. The corollary also holds when going from larger to smaller pixels.
The cost of this regridding operation is changes in any fixed-amplitude noise:
regridding larger pixels to smaller ones increases the noise per pixel,
while smaller to larger decreases the noise in the larger pixel.

This averaging property of pixels allows us to write the relation 
\begin{equation}
\label{eq:grid}
P_{j} = \frac{1}{r} \sum_{i}^{r} \tilde{P}_{i}^{k},
\end{equation} 
where $\tilde{P}^{k}$ is the image of the PSF sampled on a finer grid,
and the scale factor between the areas of pixels $k$ and $j$ is $r$.
Because the pixel size does not appear in Equation \ref{eq:psfstack}, we
can write
\begin{equation}
\label{eq:psfstackprime}
\tilde{P} = \sum_{i}^{N_{j}} \frac{\tilde{M}(x_{i},y_{i})}{F_{i}},
\end{equation}
where $\tilde{M}$ is a version of the image on the finer grid and is
related to $M$ through $M_{j} = (1/r) \sum_{i}^{r} \tilde{M}_{i}^{k}$. Any $r$ such that the $r$-times finer image has significantly higher resolution than the native image resolution will increase the sampling rate and facilitate the stacking method; increasing $r$ increases the small-scale structure it would be possible to capture in the reconstructed PSF. However, increasing $r$ can also lower the S/N or introduce greater numerical instability due to the larger number of subpixels per pixel. 
The fundamental cost of this superresolution stacking is that the noise per pixel is increased over the native pixel resolution. This can easily be addressed by making the list $\alpha$ large to compensate. As regridding conserves S/N in an area, the total S/N on the PSF will remain fixed for a fixed noise and number of sources in the stack.

To implement the proposed method in an unbiased manner, we
require several conditions to be true:
\begin{enumerate}
\item The source positions are uncorrelated and sources are uncrowded, preventing combined or overlapping sources from coherently adding in the stack.
\item The source list we stack on comprises unresolved sources so that
 we are reconstructing an image of the PSF.
\item The source images do not suffer from detection artifacts like
 saturation or nonlinearity.
\item $\langle \Delta \rangle = 0$, meaning that stacking over many
 sources averages down the noise.
\end{enumerate}
In practice, we can meet these requirements by making suitable choices
for the list of sources to stack on and imposing some weak
assumptions about the size of the PSF relative to the pixels. Requirements 1 and 2 are met by using a catalog of
stars on which to stack. Stars are unresolved by all but very specialized telescopes and (at middle and high galactic latitudes where source density is low) have uncorrelated positions \citep{Zemcov2014}. Requirements 3
and 4 can be met by selecting sources with fluxes faint enough to keep
in the linear regime of the detector and restricting the catalog to a
relatively narrow range of fluxes so the denominator $F_{i}$ in
Equation \ref{eq:psfstackprime} does not strongly overweight noisy sources.
However, the noise in the final PSF measurement is a function of the
number of sources in the list $\alpha$, so the tradeoff between the flux
range and the number of sources depends on the details of both the instrument and the corresponding survey.

In order to approach this problem in a computationally efficient manner, 
we stack a tractable number of small ``thumbnails'' around each source centered on
$(x_{i},y_{i})$, which focuses attention on the regions of interest
and removes the necessity of generating many shifted versions of $M$. The size of the thumbnail
image is determined by including a region large enough that a large fraction of the response from the PSF is included, without including unnecessary background noise.

\begin{figure*}[htbp!]
\centering
\includegraphics[width=6.5in]{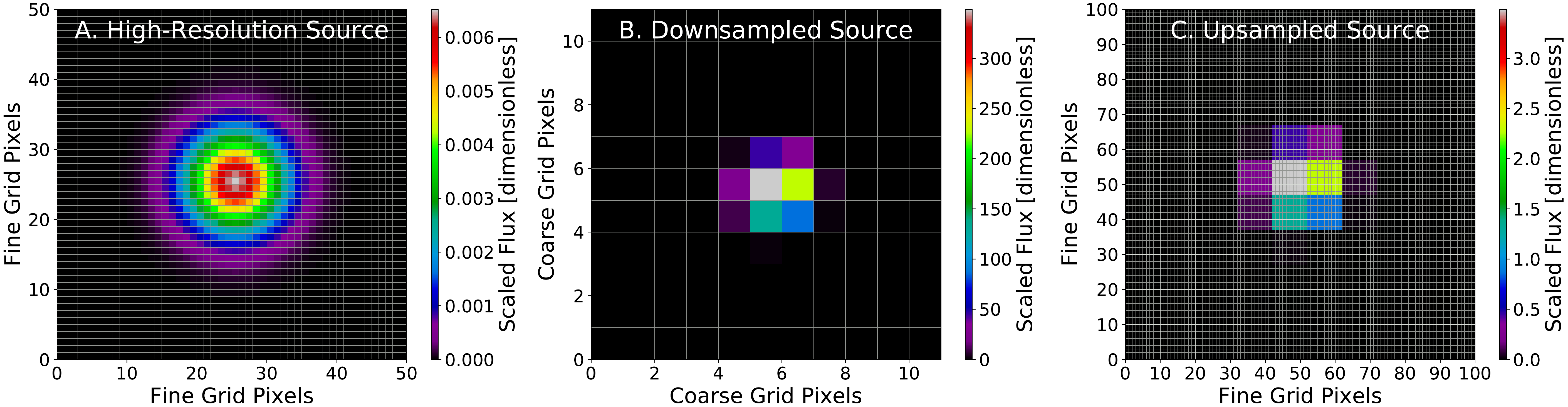}
\caption{Regridding of a single source to alter resolution. Panel A shows a single Gaussian point source generated in the fine grid. In panel B, the source resolution is sampled down by a factor of 10 to represent a simulated undersampled image (see Section \ref{S:im_sim} for details), and in panel C, the resolution is sampled back up by the same factor. Although the source contains the same number of coarse pixels in panels B and C, the source in panel C contains 100 subpixels in each of these. In panel C, the center of the source appears to have shifted from the coarse grid lines due to the subpixel center now being located with more accuracy than is possible in the coarse grid. Stacking on each source's subpixel center makes superresolution PSF reconstruction possible by allowing the stacked subpixels to generate high-resolution structure.\label{fig:regrid} }
\end{figure*}

With these requirements in mind, the stacking PSF estimation algorithm
can be broken into five steps:
\begin{enumerate}
\item Resample the sky image $M$ into an image $\tilde{M}$ on a pixel
 gridding $r$-times finer. The image will not appear different because
 a single pixel value from $M$ will fall into multiple pixels in
 $\tilde{M}$, but $\tilde{M}$ will have $r$-times more pixels on a
 side than $M$ (see Figure \ref{fig:regrid}).

\item For each star in the list $\alpha$, cut out a thumbnail $T$ centered on
 the known position of a source $(x,y)$. The size of the thumbnail
 depends on the desired angular extent of the final PSF estimate. Due to the nature of resampling, the source's subpixel center will not necessarily be centered in a coarse-grid pixel. The thumbnail is cut such that the subpixel center will be aligned with the center of the stack.

\item Suppress the overall constant offset value of the thumbnail. For equally bright
 sources, subtracting the mean of the thumbnail image is acceptable,
 but for sources of varying brightness, an estimate of the sky
 brightness away from the star is a good choice.

\item Add the thumbnail centered on the known source position into the
 stack. 

\item Repeat for all stars in the catalog. All of the source thumbnails are combined by taking the mean of all corresponding pixels (i.e., the $i$th pixel in the stack contains the mean of all the $i$th pixels in each thumbnail), and this becomes the stacked PSF, $P_{\rm S}$. This is a superresolution image of the underlying optical PSF, $P_{\rm true}$, convolved with the pixel grid function, $P_{\rm grid}$. Deconvolution is still necessary to return $P_{\rm true}$ \citep{guillard}.
\end{enumerate}

\begin{deluxetable*}{ccccc}[htb!]
\tablecaption{Source Crowding as a Function of Galactic Latitude \label{table:crowdtab}}
\tablehead{\colhead{} & \colhead{\bf{Total Available Sources}} & \colhead{\bf{Isolated }} & \colhead{\bf{Sources}} & \colhead{\bf{Total Sources}} \\
\colhead{\boldmath{$(\ell,b)$}} & \colhead{\bf{with 11 \boldmath{$\leq$ $m_{\rm AB}$ $\leq$} 15}} & \colhead{\bf{Sources}} & \colhead{\bf{from Masking}} & \colhead{\bf{in Stack}}}
\startdata
 (0\degree, 90\degree) & 1653 & 540 & 808 & 1348 \\
 (0\degree, 60\degree) & 2078 & 438 & 1169 & 1607\\
 (0\degree, 30\degree) & 4710 & 48 & 2292 & 2340\\
 (0\degree, 15\degree) & 6794 & 5 & 2185 & 2190\\
\enddata
\end{deluxetable*}

Source crowding is a concern when stacking sources. Sources appearing in a thumbnail that have brightness similar to the stacking target will contribute flux to the stack and broaden the estimate of $P$. In the limit that source positions are uncorrelated, interloper sources have random positions and so act as an extra source of noise in $P_{\rm S}$. However, with a finite number of sources there may be significant sample variance from stack to stack. To mitigate this problem, we mask interloper sources as part of the stacking procedure by measuring the distance between the target and any interlopers. Interlopers with center position $\leq 11$ coarse-grid pixels from the target source's center are masked by excluding the thumbnail pixels within a radius of five pixels from the interloper position from the sum. If the two sources are closer than 8 coarse-grid pixels apart and the target source is not at least an order of magnitude brighter than the interloper, that source is rejected from the stack entirely. Table \ref{table:crowdtab} demonstrates the effects of our crowding cut on representative $3.5 \times 3.5$ square-degree fields at various galactic latitudes with fixed longitude. 
As an example, at ($\ell$, $b$) = (0\degree, 90\degree) 18\% of sources are rejected, while in a crowded field at ($\ell$, $b$) = (0\degree, 15\degree) 68\% of sources are rejected. A benefit of the algorithm discussed here is that, even in such crowded fields, we are able to reconstruct a useful estimate for $P_{\rm true}$ (see Section \ref{S:sphx}).

\subsection{Reconstructed PSF Deconvolution}
\label{sec:decon}

\begin{figure*}[htbp!]
\centering
\includegraphics[width=5in]{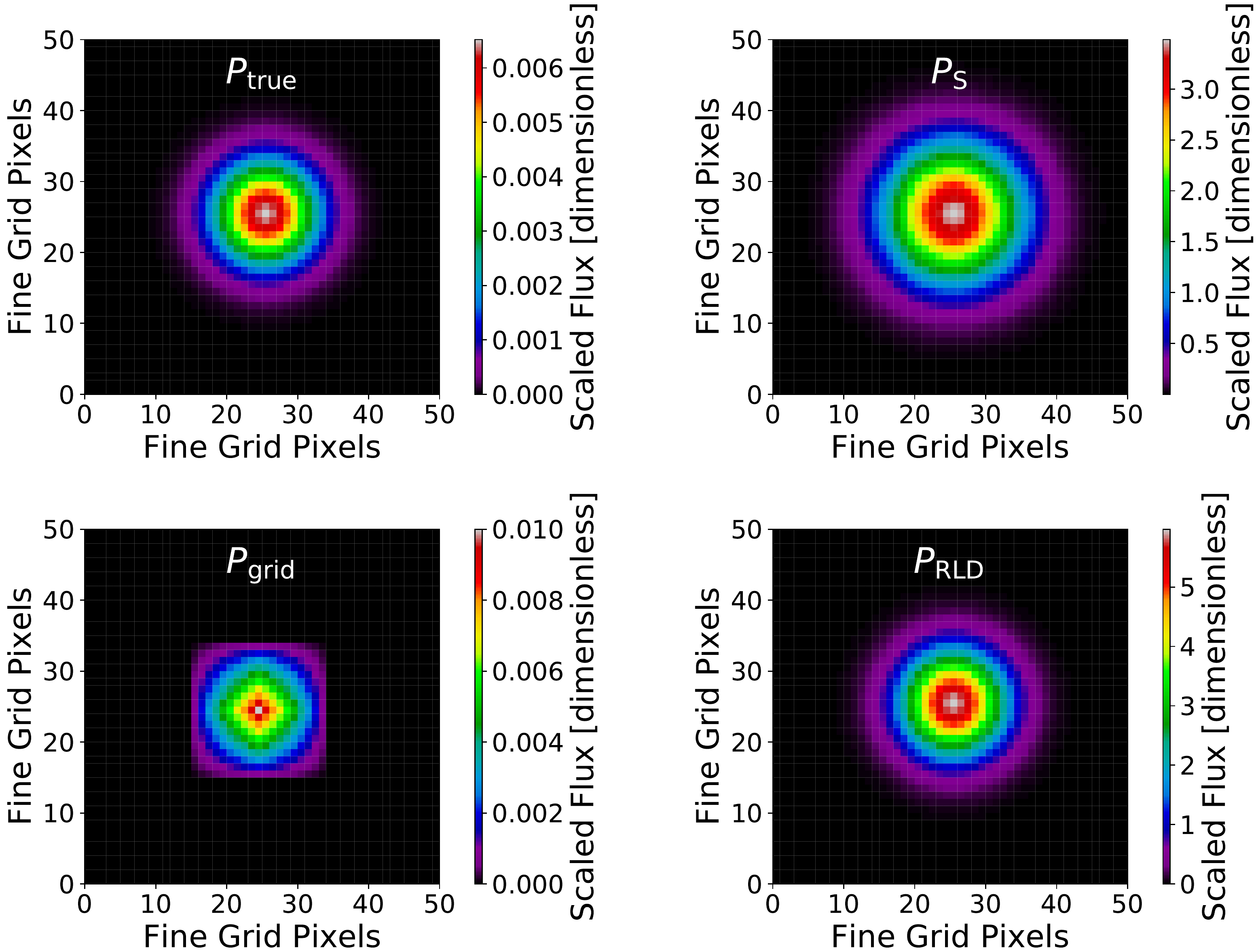}
\caption{Deconvolution of the pixel grid function. In the top left, $P_{\rm true}$ for a Gaussian source is shown, followed by $P_{\rm S}$ (top right) for a stack of 1000 Gaussian sources at a scaling factor of $r$ = 10, $P_{\rm grid}$ (bottom left) that is computed using Equation \ref{eq:psfpix} and deconvolved from $P_{\rm S}$ via RLD, and the resulting $P_{\rm RLD}$ (bottom right). $P_{\rm RLD}$ exactly matches $P_{\rm true}$, further demonstrated in Figure \ref{fig:radstack}. \label{fig:pgrid}}
\end{figure*}

Due to the oversampling inherent to our superresolution stacking method, we require a deconvolution step in order to remove $P_{\rm grid}$ and return $P_{\rm true}$ from $P_{\rm S}$ \citep{Lauer1999,decon_review,SR2003}. This type of issue has also arisen in Planck data, for which the optical beam cannot be recovered without the deconvolution of the effect of sampling time \citep{planck_beams,planck_char}. \cite{guillard} follow a similar procedure of superresolution PSF reconstruction followed by deconvolution for the Mid-Infrared Instrument on the James Webb Space Telescope, although their superresolution method combines multiple images followed by deconvolution via a maximum a posteriori method.
We choose the Richardson--Lucy deconvolution (RLD), a common algorithm used in this type of problem \citep{Richardson1972,Lucy1974}. This algorithm is also referred to as the expectation-maximization method and is a form of maximum likelihood estimation \citep{decon_review}. We have implemented RLD on $P_{\rm S}$ according to the following prescription.
The known blurring factor $P_{\rm grid}$ (shown in Figure \ref{fig:pgrid}) is a matrix where each pixel contains the fraction of light detected from a point source:
\begin{equation}
\label{eq:psfpix}
{P_{\rm grid}(x,y) = [r - (x-x_{0})][r - (y-y_{0})]}
\end{equation}
where $r$ is the upscaling factor, and $x_{0}$ and $y_{0}$ are the center of the stacking area. 
Starting from $P_{\rm S}$ and $P_{\rm grid}$, each iteration of the RLD proceeds as follows:
\begin{equation}
\label{eq:rld}
{P_{\rm RLD}^{i+1} = P_{\rm RLD}^{i} \bigg(\frac{P_{\rm S}}{P_{\rm RLD}^{i}\ast{}P_{\rm grid}}\ast{}P_{\rm grid}^{\rm ref}\bigg)},
\end{equation}
where the initial $P_{\rm RLD}^{i}$ is simply $P_{\rm S}$, and $i$ increases on every iteration for $I_{\rm RLD}$ total iterations. $P_{\rm grid}^{\rm ref}$ is the reflection of $P_{\rm grid}$ across both axes. The final result is $P_{\rm RLD}$, the deconvolution of $P_{\rm S}$ and a more accurate reconstruction of $P_{\rm true}$. This process is demonstrated for ideal Gaussian sources in Figure \ref{fig:pgrid}. However, we find that there is a limit to the resulting improvement in reconstruction quality available from increasing the number of RLD iterations due to RLD's tendency to amplify artificial structure after too many iterations \citep{hanisch_white_gilliland_1997}. This is problematic for applications that require fine features of the PSF, for example, optimally weighted photometry (see Section \ref{sec:optphot}). 

In order to improve the quality of the reconstructed PSF beyond that which can be achieved by RLD, we implement a more advanced method of deconvolution with the iterative back-projection (IBP) algorithm, in which the error or difference between simulated and observed low-resolution images is iteratively reduced \citep{IraniPeleg1991}. This method is also based on maximum likelihood estimation but was developed for superresolution image reconstruction. The approach is similar to back-projection used in tomography and has previously been used in reconstruction of blurred or degraded images. It is similar to RLD in that it offers no unique solution and has the potential to falsely amplify noise after too many iterations \citep{SR2003}. This method has been applied in remote sensing and planetary science \citep{ibp_mars}, and other versions of this method have been used to analyze images of solar flares \citep{bp_solar} and in the testing of superresolution image reconstruction methods on simulated Euclid data \citep{sr_euclid}. 

\begin{figure*}[ht]
\centering
\includegraphics[width=6.5in]{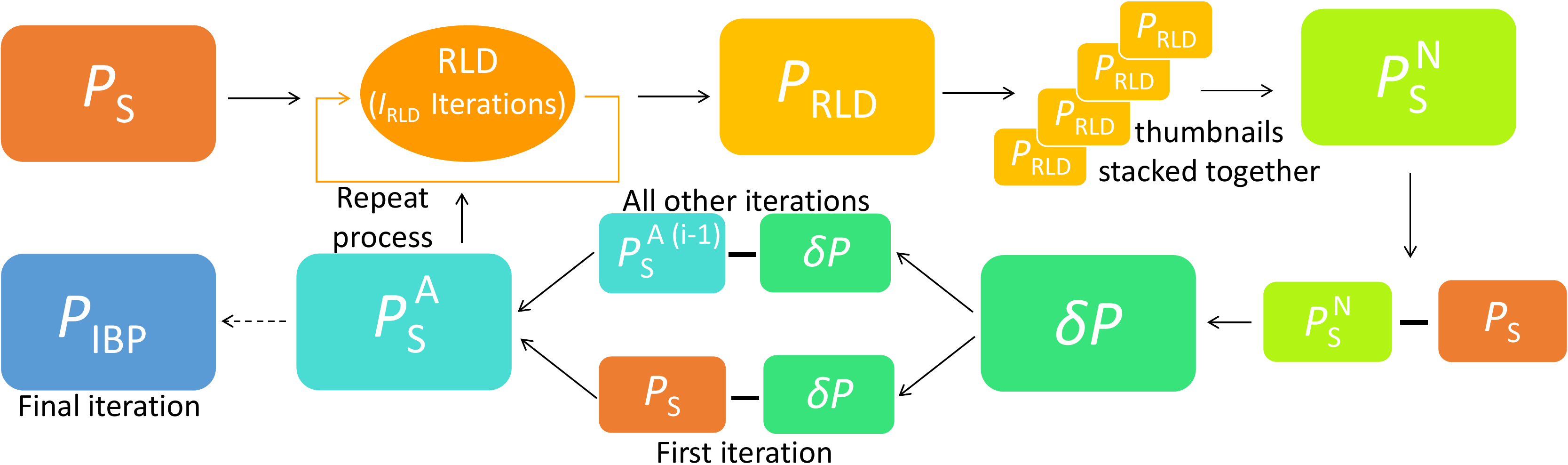}
\caption{Steps of the iterative back-projection algorithm. On the first iteration $P_{\rm S}$ is deconvolved using the regular RLD method, while on all subsequent iterations $P_{\rm S}^{\rm A}$ is deconvolved. Deconvolution is followed by a repeat of the stacking procedure using $P_{\rm RLD}$ as $P_{\rm true}$, creating $P_{\rm S}^{\rm N}$. The error is calculated to be the difference between $P_{\rm S}$ and $P_{\rm S}^{\rm N}$, and that error is subtracted from either $P_{\rm S}$ (if first iteration) or the previous $P_{\rm S}^{\rm A}$, creating the new $P_{\rm S}^{\rm A}$. The cycle then repeats with deconvolution via RLD again, resulting in $P_{\rm IBP}$ on the final iteration. \label{fig:flowchart}}
\end{figure*}

After $P_{\rm S}$ is obtained, the IBP iterations proceed as follows:
\begin{enumerate}
\item An RLD is performed as previously described, with $I_{\rm RLD}$ iterations. This produces $P_{\rm RLD}$.
\item The entire stacking procedure is repeated using $P_{\rm RLD}$ as $P_{\rm true}$. $P_{\rm RLD}$ is placed into the fine grid (using the same number of sources as were used previously), and the fine grid is downsampled and upsampled by the previously used scale factors. These sources are stacked as before to produce the new stacked PSF, $P_{\rm S}^{\rm N}$. 

\item An error term $\delta P$ is calculated as
\begin{equation}
\label{eq:ibperr}
{\delta P= P_{\rm S}^{\rm N} - P_{\rm S}}.
\end{equation}

\item $\delta P$ is used to produce an error-adjusted stack, $P_{\rm S}^{\rm A}$. On the first iteration, $P_{\rm S}^{\rm A}$ is created via
\begin{equation}
\label{eq:adjstack}
{P_{\rm S}^{\rm A} = P_{\rm S} - \delta P}.
\end{equation}
On every subsequent iteration, $P_{\rm S}^{\rm A}$ is
\begin{equation}
\label{eq:adjstack2}
{P_{\rm S}^{\rm{A} \it{(i+\rm{1})}} = P_{\rm S}^{\rm{A} \it{(i)}} - \delta P}.
\end{equation}

\item The cycle repeats with the RLD of the newly formed $P_{\rm S}^{\rm{A} \it{(i+\rm{1})}}$ (which is resampled and stacked to form $P_{\rm S}^{\rm N}$) until a specified completion criterion is reached. The result of the final iteration is $P_{\rm IBP}$, which is the updated and more accurate reconstruction.
\end{enumerate}
The overall flow of this algorithm is illustrated in Figure \ref{fig:flowchart}.

Standard RLD fails in the presence of noise for several reasons, including the amplification of noise over many iterations and, depending on the choice of image zero point, the presence of negative values. This arises because the RLD algorithm is based on the maximum likelihood for Poisson statistics, so its solution requires positive input data. However, stacking requires a choice about the image zero point assumed in each thumbnail that may produce negative values into $P_{\rm S}$. Though it is possible to design algorithms that do not have this feature, the most general case is that we have to account for mildly negative values in $P_{\rm S}$. 

We handle this problem by implementing a more advanced version of the RLD algorithm \citep{hanisch_white_gilliland_1997} that adds an initial estimate of the background and read-noise value to every pixel and suppresses the contribution from pixels with a value less than the damping factor in each iteration. This damping prevents the amplification of noisy pixels that do not contain structure related to the PSF. Because our PSF is concentrated in the center of the stacked image, we use a damping matrix that performs no damping in the center but damps heavily beyond a circular radius of 13 fine-grid pixels from the center, preventing each iteration from drastically changing the values of damped pixels. Beyond a 13 pixel radius, the PSF is largely noise dominated and does not contribute much signal to the optimal photometry. As presented below, this version of the IBP algorithm successfully deconvolves $P_{\rm grid}$ from $P_{\rm S}$ even in the presence of noise. 

\subsection{Using a Reconstructed PSF for Optimal Photometry}
\label{sec:optphot}

Optimal photometry \citep{Naylor1998} is an example of an application in which detailed knowledge of the PSF is required to reach the maximum possible S/N on point source fluxes. The best estimate for $P_{\rm true}$ (here $P_{\rm IBP}$ unless explicitly noted) is shifted via interpolation to account for the difference between the source's known coordinates via catalog reference and those same coordinates scaled by the regridding factor, $r$. It is then downsampled by $r$ to match the resolution of the source. This shifted and resampled reconstructed PSF is normalized and used to give each pixel the weight of its individual flux contribution. This weight is defined as
\begin{equation}
\label{eq:photweight}
{W_{\rm P}} = \frac{P_{\rm P}}{\sum ({P_{\rm P}}^2)},
\end{equation}
where $P_{\rm P}$ is the shifted and resampled reconstructed PSF. Each source's flux is then computed as
\begin{equation}
\label{eq:weightflux}
{F_{\rm P}} = \sum (T\cdot W_{\rm P}),
\end{equation}
where $T$ is a thumbnail cutout of each source, and $W_{\rm P}$ is the previously calculated weight function. For an image with many point sources, we define the average deviation of all source fluxes determined via optimal photometry from known (catalog) values to be
\begin{equation}
\label{eq:fdev}
{\langle \delta F \rangle} = \frac{\sum\limits_{i}^{N} [(F_{\rm{P},\it{i}}-F_{i})/F_{i}]}{N},
\end{equation}
where $\langle \delta F \rangle$ is expressed in percent, $N$ is the number of sources in the image, $F$ is a source's known flux, and $F_{\rm P}$ is that computed by optimal photometry. 

\subsection{Image Simulation}
\label{S:im_sim}

In order to test and characterize these algorithms, simulated images of the sky are necessary. We simulate images of undersampled point sources consistent with the SPHEREx expectation \citep{Korngut2018} according to the following prescription. First, we generate a grid that represents the $r$-times finer grid or pixelized image. The native detector size for SPHEREx will be 2,048 $\times$ 2,048 pixels covering a $3.5 \times 3.5$ deg$^{2}$ field of view (FoV) in each of the six main wavelength bands. 
We select $r$ to be 10 for Gaussian $P_{\rm true}$, so that our $r$-times finer image contains 20,480 $\times$ 20,480 pixels, or $r$ to be 20 for SPHEREx $P_{\rm true}$, so that the fine-grid image contains 40,960 $\times$ 40,960 pixels. This increases the sampling rate for SPHEREx to $\sim$6 pixels per FWHM, which allows for superresolution recovery of the PSF. The parameter $r$ that we use to simulate the high-resolution image does not need to have the same value as that used to increase resolution during stacking. We choose to set them equal for simplicity, but for real data, the stacking $r$ can be empirically determined. We populate the generated image with point sources using one of two methods:
 
 \begin{enumerate}
   \item One thousand point sources are randomly placed over the FoV, and for simplicity, all sources are given uniform flux $F_{\nu}$ as a dimensionless quantity such that the S/N is chosen to be 20. An S/N of 20 corresponds to $m_{\rm AB}$ = 17.87. See the Appendix for more details on the units we use to define flux and how we calculate S/N.
   \item Realistic star fields are generated for any desired sky position with an all-sky catalog of 332 million sources, derived by selecting stars in the Gaia DR2 \citep{gaia} catalog with close counterparts (within 1$^{\prime \prime}$ angular separation) in the AllWISE catalog \citep{allwise}. The use of the $g$ and $r_p$ photometry from Gaia in combination with the W1, W2, and W3 photometry from WISE gives a rough SED for each source from which realistic fluxes for the SPHEREx bands can be estimated. 
   For the special case of single-field tests presented in Section \ref{S:results}, we use the field centered at the north galactic pole (NGP; ($\ell$, $b$) = (0\degree, 90\degree)), where star coordinates are uncorrelated and fields are the least crowded. This ``minimal'' field contains $\sim$20,000 sources ranging from $3 < m_{\rm AB} < 21$. 
 \end{enumerate}

After the image of point sources gridded according to $P_{\rm grid}$ is created, the map is convolved with the $P_{\rm true}$ (either Gaussian or SPHEREx\footnote{The SPHEREx $P_{\rm true}$ is derived from optical simulations performed by L3-SSG as part of the SPHEREx Phase A study using an end-to-end telescope design.}) under study. This image is then sampled down by $r = 10$ or $r = 20$ as appropriate to the native image resolution. Figure \ref{fig:regrid} demonstrates how the regridding process works for a single source.

\section{Results}
\label{S:results}

\begin{deluxetable}{cccccc}[htbp!]
\tabletypesize{\scriptsize}
\tablecaption{Summary of Simulations\label{table:simtab}}
\tablehead{\colhead{\bf{PSF Type}} & \colhead{\bf{Flux Type}} & \colhead{\bf{RLD/IBP}} & \colhead{\bf{Noise}} & \colhead{\bf{Figures}} & \colhead{\bf{Sections}}}
\startdata
 Gaussian & Uniform & RLD & No & \ref{fig:radstack} & \ref{S:gauss_uniform} \\
 Gaussian & Catalog & RLD & No & \ref{fig:sc_gauss_flux} & \ref{S:gaia} \\
 Gaussian & Catalog & IBP & No & \ref{fig:gauss_gaia_ibp} & \ref{S:gaia} \\
 Gaussian & Catalog & IBP & Yes & \ref{fig:gauss_noise} & \ref{S:in_noise} \\
 SPHEREx & Catalog & RLD & No & \ref{fig:ibpdecon},\ref{fig:fourier} & \ref{S:sphx} \\
 SPHEREx & Catalog & IBP & No & \ref{fig:ibpdecon},\ref{fig:fourier},\ref{fig:ibpflux} & \ref{S:sphx}\\
 SPHEREx & Catalog & IBP & Yes & \ref{fig:noisy_decon},\ref{fig:mag1},\ref{fig:fomperc},\ref{fig:sc_noise},\ref{fig:sphx_noise} & \ref{S:sphx} 
\enddata
\end{deluxetable}

We now have all the pieces required to test the described method against various cases, including nonideal PSF shapes, noise, and crowding, and to assess its overall performance. In Sections \ref{S:gauss_uniform} -- \ref{S:in_noise}, we apply the IBP method to Gaussian PSFs in various scenarios to develop intuition. In Section \ref{S:sphx}, we introduce the SPHEREx $P_{\rm true}$ and assess the effects of noise, field position, and other quantities of interest. Table \ref{table:simtab} gives a summary of the various tests and the figures and sections where the corresponding results can be found. Finally, in Section \ref{S:lorri}, we apply the IBP PSF reconstruction to data from the LORRI instrument on New Horizons and are able to identify additional complicating factors present in real data such as pointing instability.

\subsection{Gaussian Point Sources with Uniform Flux}
\label{S:gauss_uniform}

\begin{figure*}[htb!]
\centering
\includegraphics[width=6in]{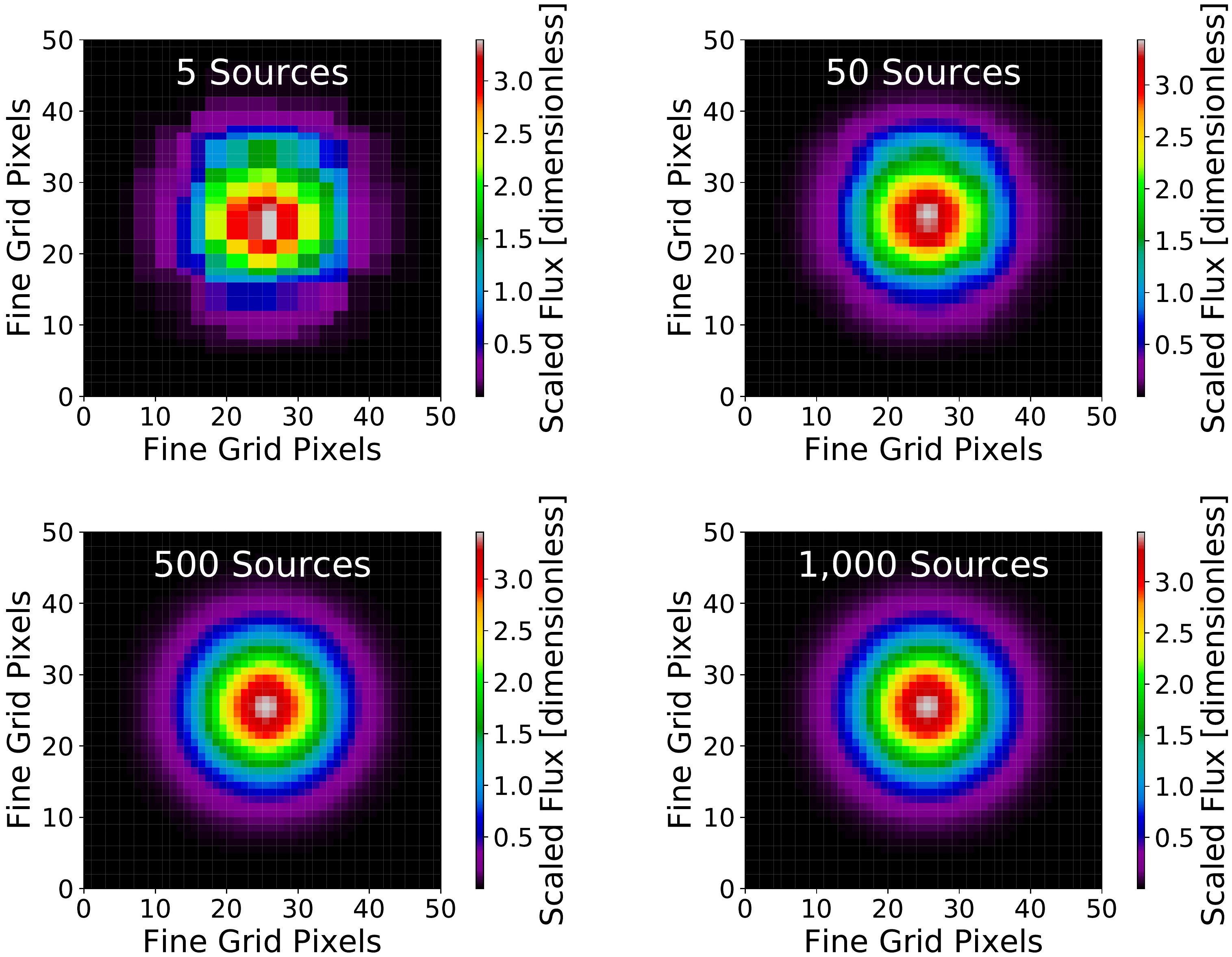}
\caption{Demonstration of stacking an increasing number of sources. On the top left, five source thumbnails are combined together via their mean. Each new thumbnail is centered on the corresponding source's known coordinates. The number of sources increases on the top right to 50 sources, on the bottom left to 500 sources, and on the bottom right to 1000 sources. Although each individual thumbnail bears little resemblance to $P_{\rm true}$, the more thumbnails that are stacked together, the more closely the stack begins to match $P_{\rm true}$.\label{fig:stacking}}
\end{figure*}

\begin{figure*}[htb!]
\centering
\includegraphics[width=6.5in]{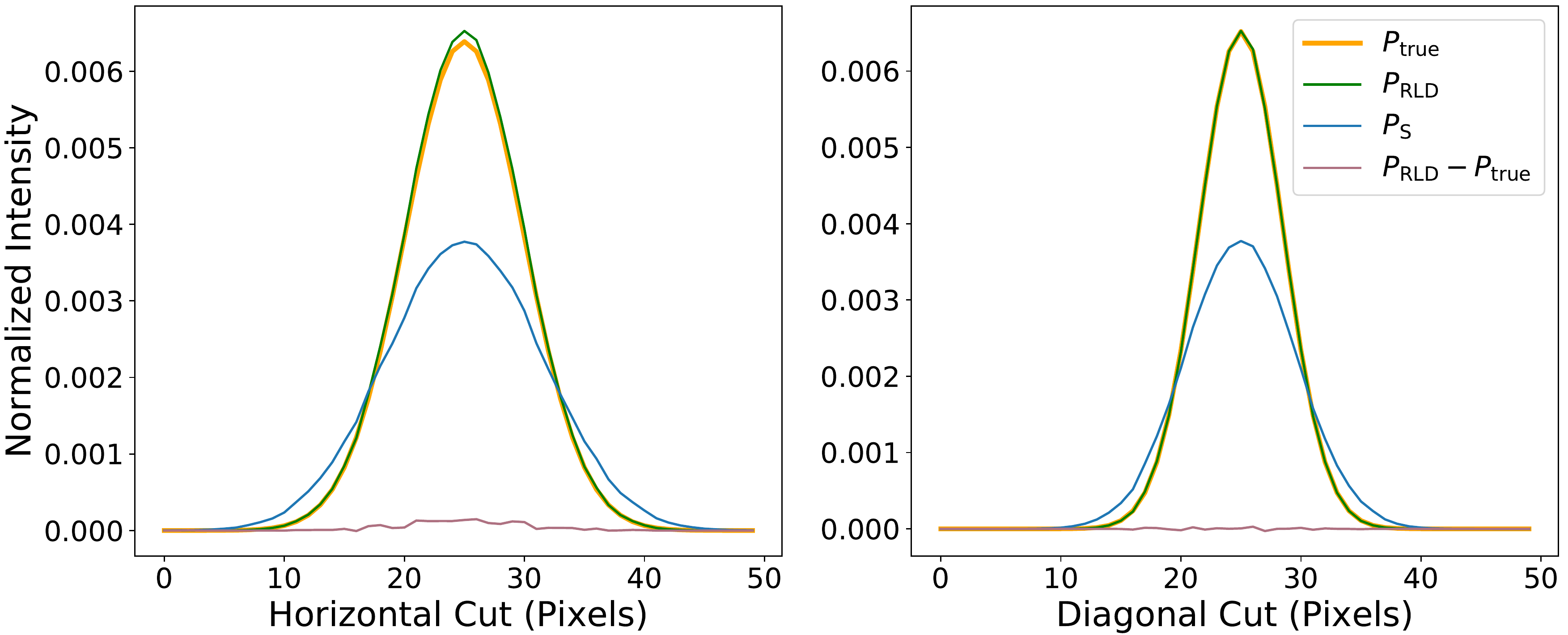}
\caption{Comparison of $P_{\rm RLD}$ to $P_{\rm S}$ and $P_{\rm true}$ for a stack of Gaussian point sources. The left panel shows a horizontal cut comparison of the various PSFs and the difference between the input PSF, $P_{\rm true}$, and the output PSF, $P_{\rm RLD}$. The horizontal cut (left panel) is a horizontal section with respect to the pixel grid through the center of the PSF, while the diagonal cut (right panel) is a section from the top-left to the bottom-right corner. While $P_{\rm S}$ is much broader than $P_{\rm true}$, the deconvolved $P_{\rm RLD}$ very closely matches the width and shape of the optical PSF. 
\label{fig:radstack}}
\end{figure*}

We first explore the fundamental properties of the PSF estimation method using a simple Gaussian model without noise. This allows us to perform an idealized test of the various stacking and deconvolution methods without complicating factors that may introduce their own sources of error. We construct a simulated image as specified in Section \ref{S:im_sim} with Gaussian point sources and uniform flux. We begin with the stacking procedure described in Section \ref{S:stack}, which is demonstrated for Gaussian point sources in Figure \ref{fig:stacking}. To evaluate the quality of $P_{\rm S}$ as a reconstruction of $P_{\rm true}$, we compare sections through $P_{\rm S}$ and $P_{\rm true}$ in Figure \ref{fig:radstack}. As expected, we find that $P_{\rm S}$ is significantly wider than $P_{\rm true}$ as $P_{\rm S}$ is $P_{\rm true}$ convolved with $P_{\rm grid}$. 

We next evaluate $P_{\rm S}$'s performance when used as the kernel for the weighted optimal photometry described in Section \ref{sec:optphot}, and find that the measured fluxes using $P_{\rm S}$ have a wide spread and are $>$20\% larger than their expected values from $P_{\rm true}$. To improve the accuracy of the photometry, we implement $I_{\rm RLD}$ = 10 iterations of the RLD algorithm. 
Figure \ref{fig:radstack} shows the horizontal and diagonal profiles of $P_{\rm RLD}$, as well as the difference between $P_{\rm RLD}$ and $P_{\rm true}$, which is negligible compared to the amplitudes of $P_{\rm S}$ and $P_{\rm true}$.

In order to quantify the accuracy of $P_{\rm RLD}$ as a pixel-weighting kernel for photometry, we simulate 50 realizations of noiseless, constant-flux sources with randomized source coordinates. We find $\langle \delta F \rangle = 0.379$\% $\pm$ $0.007$\%, which is evidence for a significant output flux bias in the most idealized constant-flux case. 

\subsection{Gaussian Point Sources with Catalog-based Flux}
\label{S:gaia}

\begin{figure*}[htb!]
\centering
\includegraphics[width=6.5in]{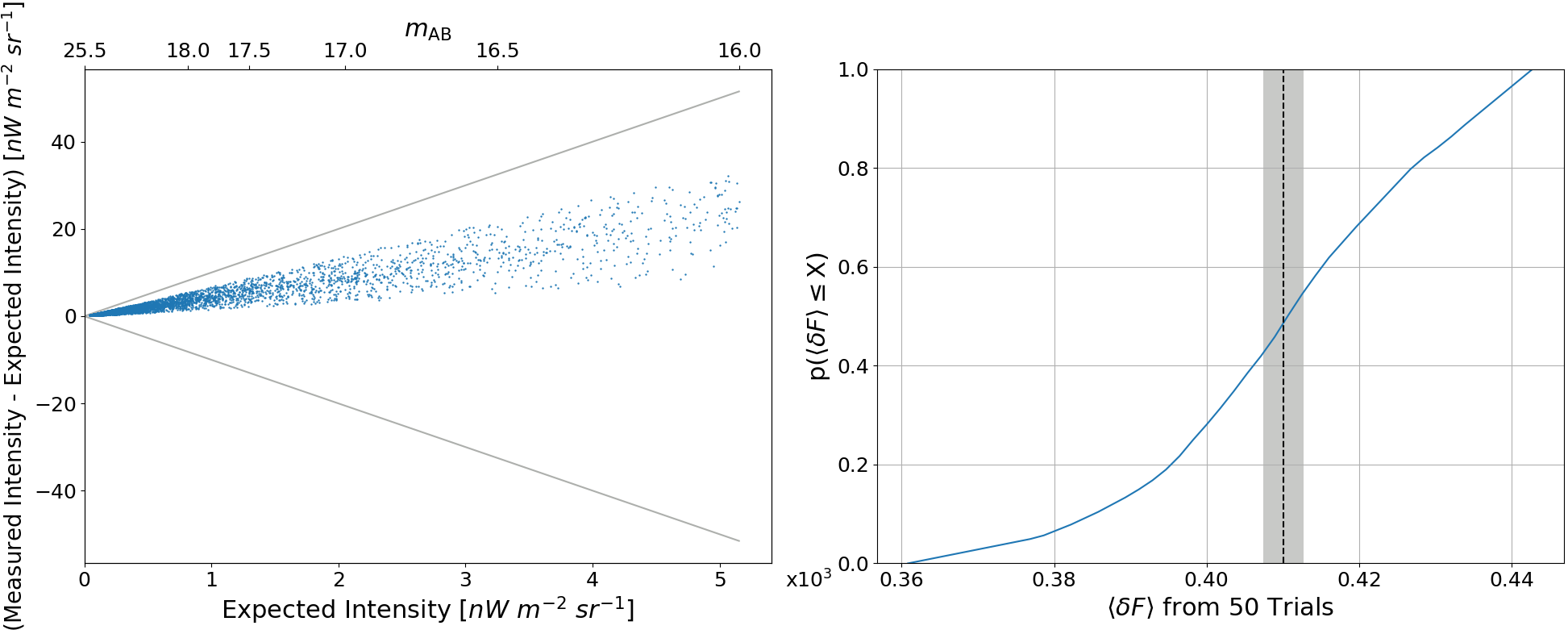}
\caption{The left panel shows the expected vs. measured flux for noiseless simulations of Gaussian PSFs in the field located at the NGP. Gray lines mark our photometric benchmark within which all fluxes have less than a 1\% difference from their known values. Photometry is only performed on sources with $m_{\rm AB} > 16$ as this marks the (approximate) upper limit of unresolved galaxy brightness for SPHEREx. Though the variation is within the requirement, a clear positive bias of the output flux can be seen. The right panel shows the cumulative distribution function (CDF) of $\langle \delta F \rangle$ for 50 trials of fields with $b > 70$\degree. The mean deviation, $0.410\% \pm 0.003$\%, is marked with the dotted line, with the gray shaded region indicating the standard error of the mean. The same positive bias as seen on the left is confirmed in an ensemble of fields and appears to be due to subtle biases in the RLD reconstruction of the PSF, indicating the need for a more accurate reconstruction of $P_{\rm true}$. \label{fig:sc_gauss_flux}}
\end{figure*}

To check the effect of a more realistic distribution of input fluxes on the RLD reconstruction, we simulate an image with Gaussian point sources and catalog fluxes for a field at the NGP constructed as described in Section \ref{S:im_sim}, again without noise in the image. Performing the same stacking and deconvolution procedure yields the results shown in Figure \ref{fig:sc_gauss_flux}. All fluxes fall within 1\% of their known values, but a positive bias at the level of $\sim$0.4\% remains. To study the variance in an ensemble of simulated images, instead of randomizing individual source coordinates, we calculate results for 50 randomly centered and independent fields with $b > 70$\degree. The $\langle \delta F \rangle$ from this ensemble is 0.410\% $\pm$ 0.003\%, which verifies the positive bias seen in the single trial at the NGP and indicates that the use of catalog-based flux instead of uniform flux has not had a negative impact. This bias is due to $P_{\rm RLD}$ still being slightly broader than $P_{\rm true}$ at the level of $\sim$1:~10$^{4}$. We conclude that in order to remove the bias, a more accurate reconstruction of $P_{\rm true}$ is necessary. 

\begin{figure*}[htb!]
\centering
\includegraphics[width=6.5in]{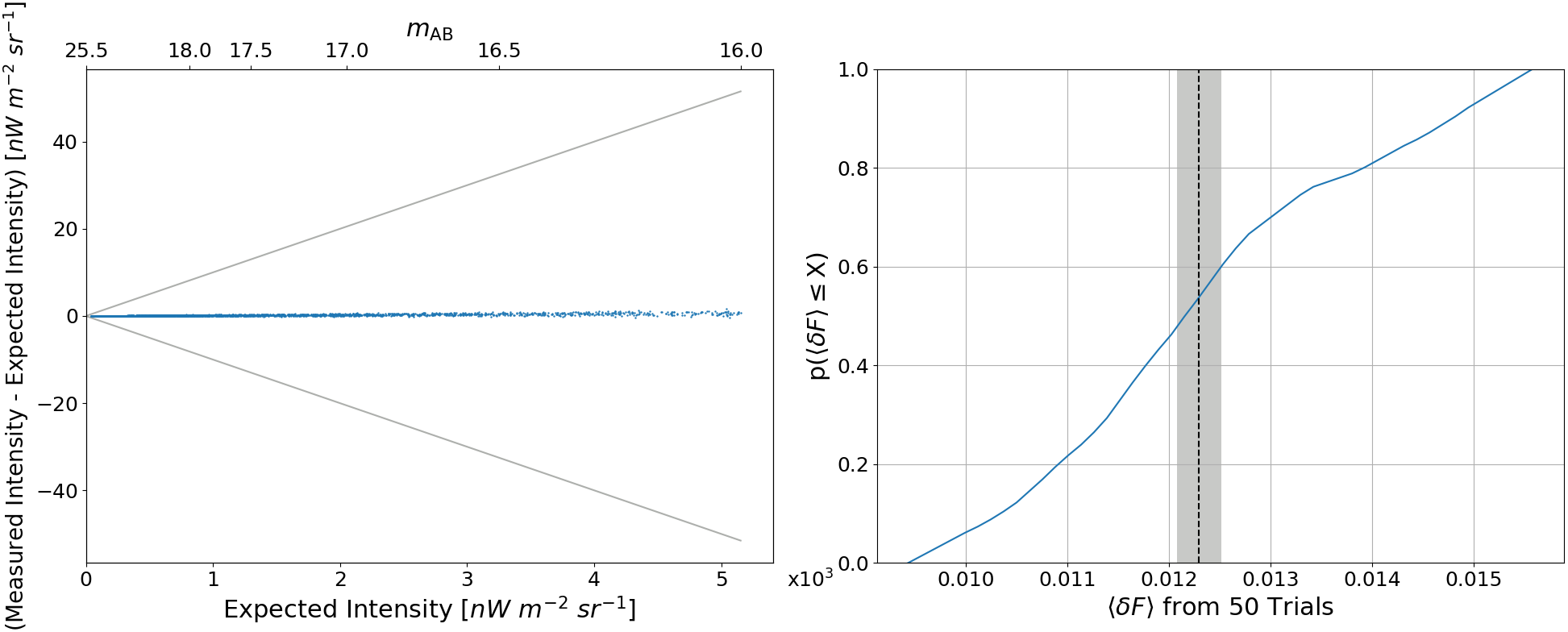}
\caption{The same test as shown in Figure \ref{fig:sc_gauss_flux} is repeated with the addition of 10 $I_{\rm IBP}$. On the left, the comparison of the difference between measured and expected flux to expected flux for a single field at the NGP reveals a much tighter correlation with bias at the $10^{-4}$ level. On the right, the CDF of $\langle \delta F \rangle$ from 50 trials with varying galactic longitude at $b > 70$\degree \ has a mean of $0.0123\% \pm 0.0002$\%, marked with the dotted line and the gray region. In this noiseless, idealized simulation IBP reduces the bias from RLD alone by an order of magnitude. \label{fig:gauss_gaia_ibp}}
\end{figure*}

In order to obtain a more accurate reconstruction, we use the IBP algorithm outlined in Section \ref{sec:decon}. Similar to $I_{\rm RLD}$, $I_{\rm IBP}$ is likewise determined by minimizing $\langle \delta F \rangle$'s total spread and bias until no further improvements can be achieved, which we empirically find converges by $I_{\rm IBP}$ = 10. Results from the more advanced algorithm for the 50 independent fields with $b > 70^{\circ}$ are shown in Figure \ref{fig:gauss_gaia_ibp} and show significant improvements in both the mean and variance of $\delta F$. This is good evidence that $P_{\rm IBP}$ is a more accurate reconstruction of $P_{\rm true}$ than $P_{\rm RLD}$. Performing the same test on sources with uniform flux also results in the previously seen bias in $\langle \delta F \rangle$ being reduced by an order of magnitude, which further demonstrates that the bias was due to reconstruction quality and not the type of source fluxes used.

\subsection{Noise}
\label{S:in_noise}

\begin{figure*}[htbp!]
\centering
\includegraphics[width=7in]{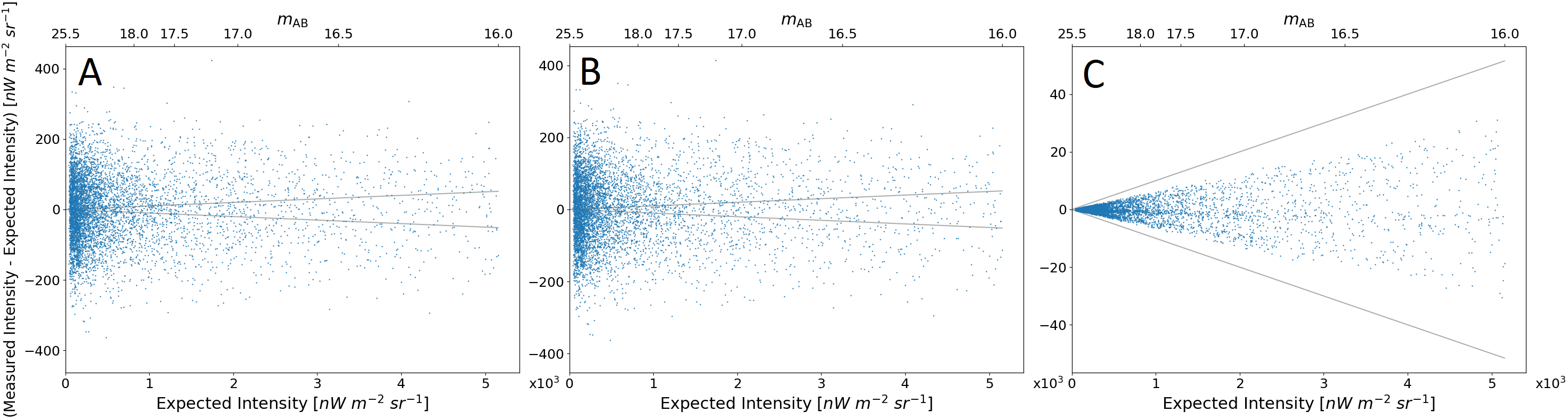}
\caption{For a single trial using a Gaussian $P_{\rm true}$ and catalog-based sources, varying photometric accuracy results depending on whether we add noise to the sources, the kernel, or both. Panel A shows the photometry using a noisy $P_{\rm IBP}$ as the kernel for noisy sources. Results are similar in panel B, which shows flux values obtained using the noiseless $P_{\rm true}$ applied to noisy sources. For both panels A and B, instrumental and photon noise dominate the photometric results. Using the noisy $P_{\rm IBP}$ as the kernel for photometry of noiseless sources (panel C) gives all flux values within the desired 1\% requirement, demonstrating that our reconstruction methods are not introducing photometric error beyond the SPHEREx requirement. \label{fig:gauss_noise}}
\end{figure*}

We simulate noise from the instrument consistent with SPHEREx Band 1 (centered at $0.93 \, \mu$m; see \citealt{spherex} for full specifications) with pixel RMS of 46 \nw\ by adding a white noise component to the native resolution source image. We also model photon noise from the sources themselves as an additional source of noise.
At this noise amplitude, a source detected at $5 \sigma$ has $m_{\rm AB} = 19.4$. We restrict the sources included in $P_{\rm S}$ to the range $10 \leq m_{AB} \leq 21.1$. The bright limit is determined by sources that would require $>$50\% correction due to nonlinearity in the SPHEREx detector, and the lower limit by sources with S/N $=1$. The latter choice is not motivated by any known limitation of the algorithm, but rather by including only those sources not dominated by noise.
In the NGP field, using $P_{\rm IBP}$ constructed from noisy sources as the kernel for optimal photometry returns photometric fluxes well within the desired 1 \% benchmark, as illustrated in Figure \ref{fig:gauss_noise}. Performing a more realistic simulation including noise in both the photometered source fluxes and the PSF stack sources results in photometry that exceeds the 1 \% requirement, but that is dominated by the noisy fluxes rather than the intrinsic error from the photometry kernel.

\subsection{Spatially Structured PSF}
\label{S:sphx}

\begin{figure*}[htb!]
\centering
\includegraphics[width=5.5in]{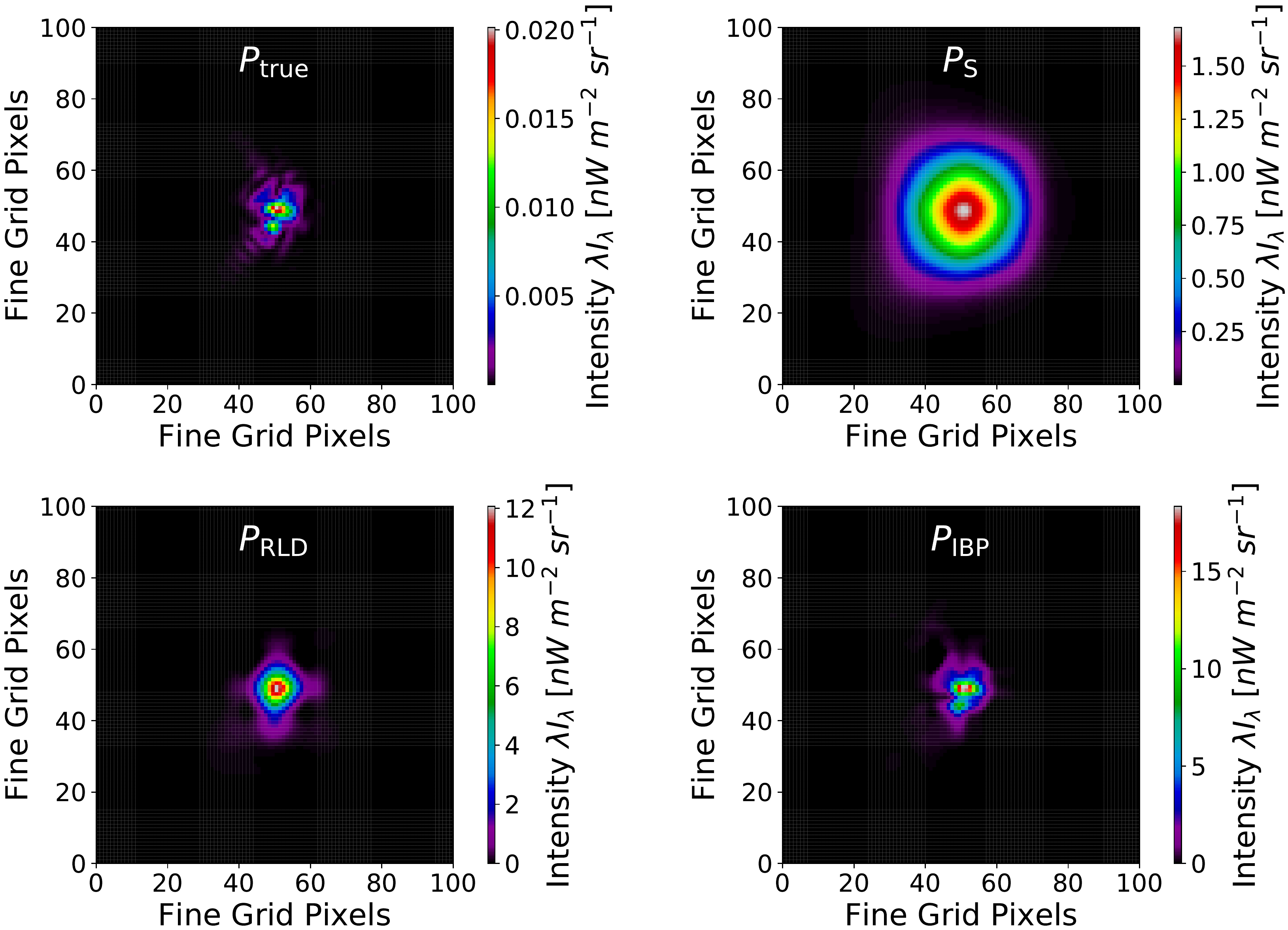}
\caption{Images of the SPHEREx $P_{\rm true}$ (top left), $P_{\rm S}$ (top right), $P_{\rm RLD}$ after 200 RLD iterations (bottom left), and $P_{\rm IBP}$ after 300 RLD iterations combined with the adaptive IBP iterations (bottom right) on $0.^{\prime \prime}31$ pixel$^{-1}$ gridding. Because the IBP offers a more accurate reconstruction of $P_{\rm true}$, it also provides more accurate photometry when used as pixel weighting. \label{fig:ibpdecon}}
\end{figure*}

In order to provide a concrete test case and assess our IBP reconstruction of the SPHEREx PSF, we simulate an image matching SPHEREx's specifications using the sky catalog for the NGP as described in Section \ref{S:im_sim}. We convolve the point sources with the SPHEREx $P_{\rm true}$ derived from optical simulations of the instrument, shown in Figure \ref{fig:ibpdecon}, and optionally add instrument and photon noise. Next, we perform the stacking procedure, pixel grid deconvolution, and optimal photometry as described in Section \ref{S:theory}. 

\begin{figure*}[htbp!]
\centering
\includegraphics[width=5.5in]{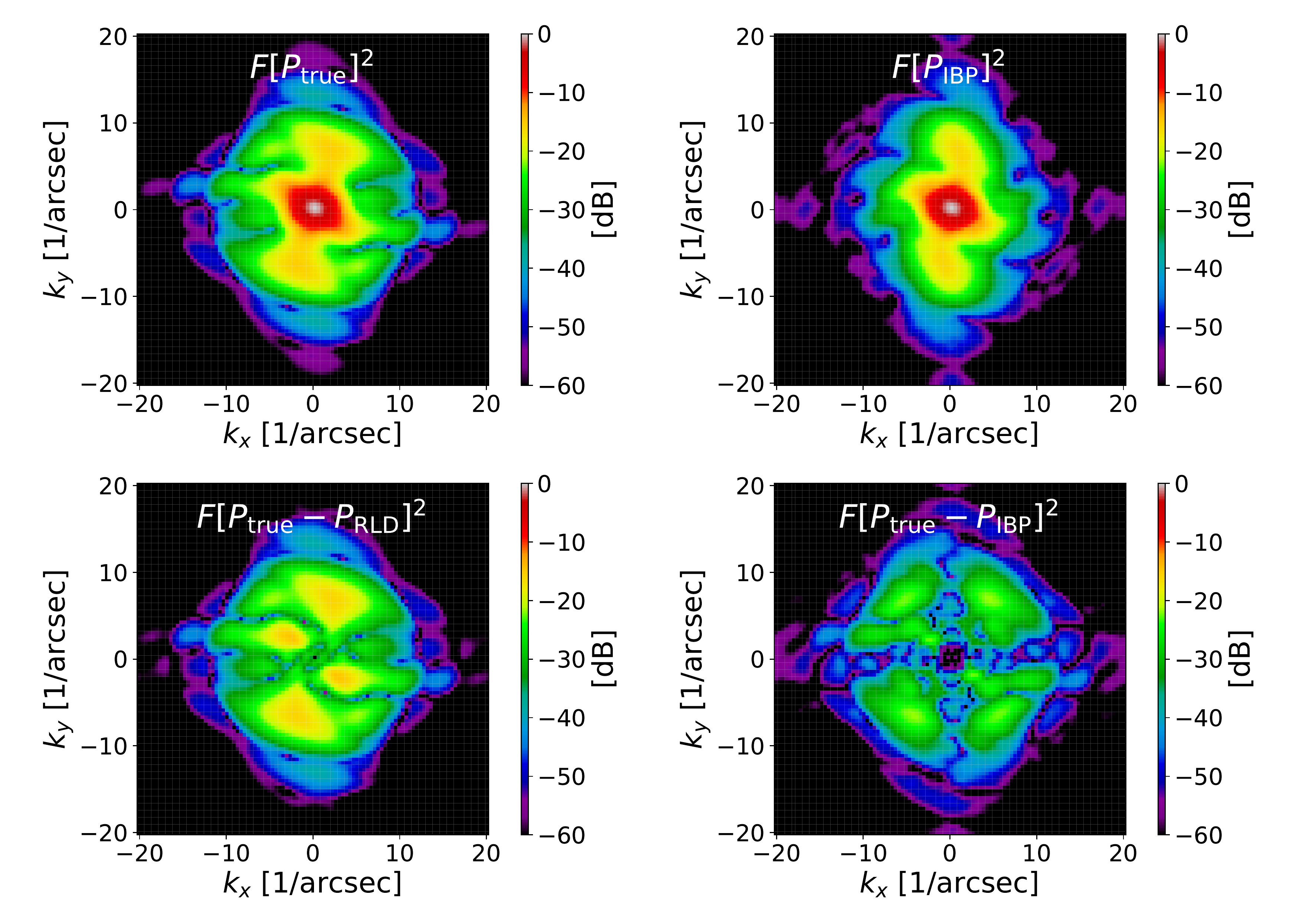}
\caption{A comparison between the squared Fourier transform of the SPHEREx $P_{\rm true}$ (top left), $P_{\rm IBP}$ (top right), difference between $P_{\rm true}$ and $P_{\rm RLD}$ (bottom left), and difference between $P_{\rm true}$ and $P_{\rm IBP}$ (bottom right). We would expect a diffraction-limited system to go to zero response near $4 \pi D/ \lambda \sim14 \,$ arcsec$^{-1}$; the measured value of $11.7 \,$ arcsec$^{-1}$ is consistent with the beam being slightly larger than diffraction limited. The IBP method exhibits residuals at the $10^{-3}$ level, which is significantly less than the traditional RLD method that shows residuals at the $> 10^{-2}$ level.\label{fig:fourier}}
\end{figure*}

\begin{figure*}[htb!]
\centering
\includegraphics[width=6.5in]{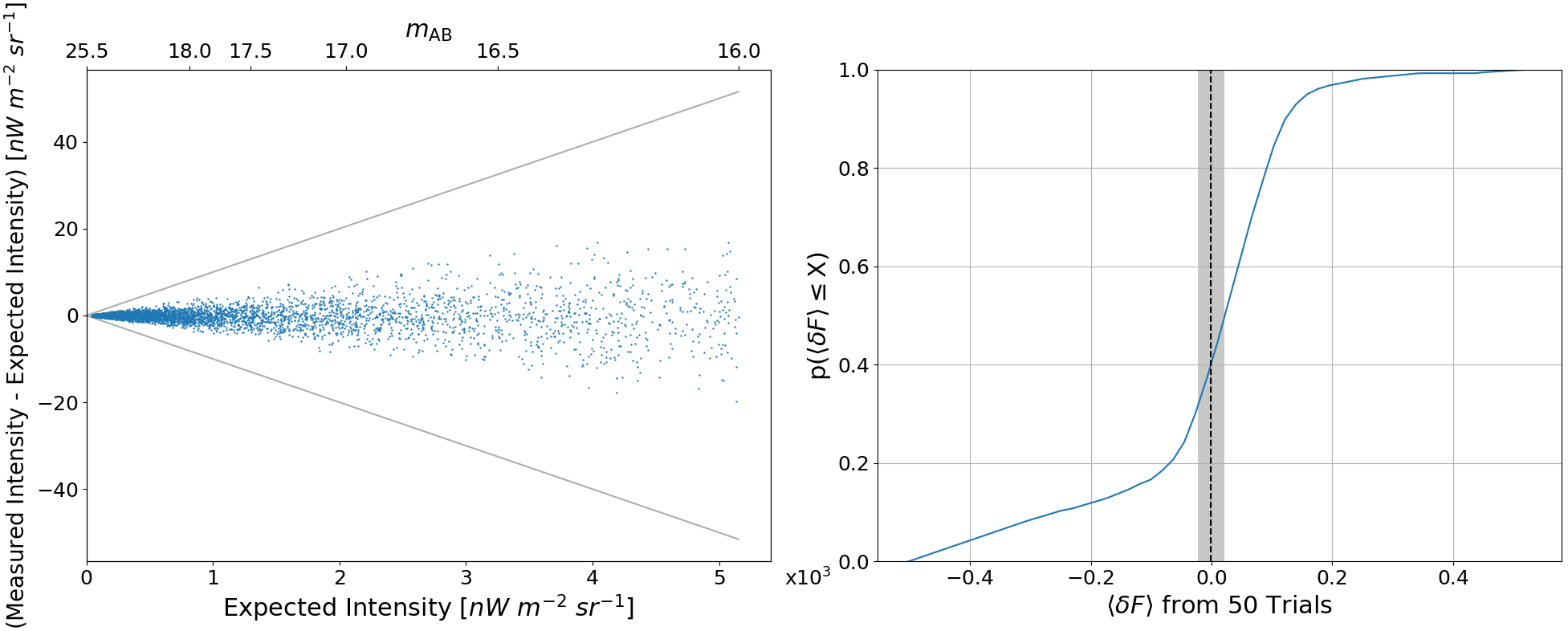}
\caption{(Left) The difference between measured and expected flux as a function of expected value, with gray lines showing the 1\% requirement for a single trial using the noiseless SPHEREx $P_{\rm true}$ and IBP algorithm. All fluxes lie within the 1\% boundaries, and we detect no overall flux bias. (Right) $\langle \delta F \rangle$ for 50 trials of fields at $b > 70$\degree \ and randomized galactic longitude, which has a mean of $-0.001\% \pm 0.022$\%. There is no statistical bias in the output fluxes from this method in this test. \label{fig:ibpflux}}
\end{figure*}

First, we perform a noiseless test. A comparison of $P_{\rm true}$ to the resulting $P_{\rm S}$, $P_{\rm RLD}$, and $P_{\rm IBP}$ calculated in the absence of noise is displayed in Figure \ref{fig:ibpdecon}. As expected from the Gaussian simulations, $P_{\rm S}$ is not similar to $P_{\rm true}$, but $P_{\rm RLD}$ with 200 $I_{\rm RLD}$ offers a significant improvement in reconstruction quality. Using the full IBP algorithm allows an even more accurate reconstruction of $P_{\rm true}$. A comparison of the squared Fourier representations of $P_{\rm true}$ and $P_{\rm IBP}$ is shown in Figure \ref{fig:fourier}, along with the corresponding differences between $P_{\rm true}$, $P_{\rm RLD}$, and $P_{\rm IBP}$. Our choice of $r$ dictates the frequency of information we are able to recover in the reconstruction. We find that using $r = 10$ for the SPHEREx $P_{\rm true}$ does not return enough high-frequency information in the reconstruction, so we use $r = 20$ to obtain accurate results. This is an empirical choice based on reconstruction quality as measured by the output photometric accuracy. Applying $P_{\rm IBP}$ as the photometry kernel results in no detectable bias within uncertainties, as demonstrated in Figure \ref{fig:ibpflux}. With this procedure, all measured fluxes are within 1\% of their known values, and 50 trials of fields with randomized galactic longitude at $b > 70\degree$ produce $\langle \delta F \rangle$ consistent with zero.

\begin{figure*}[htb!]
\centering
\includegraphics[width=5in]{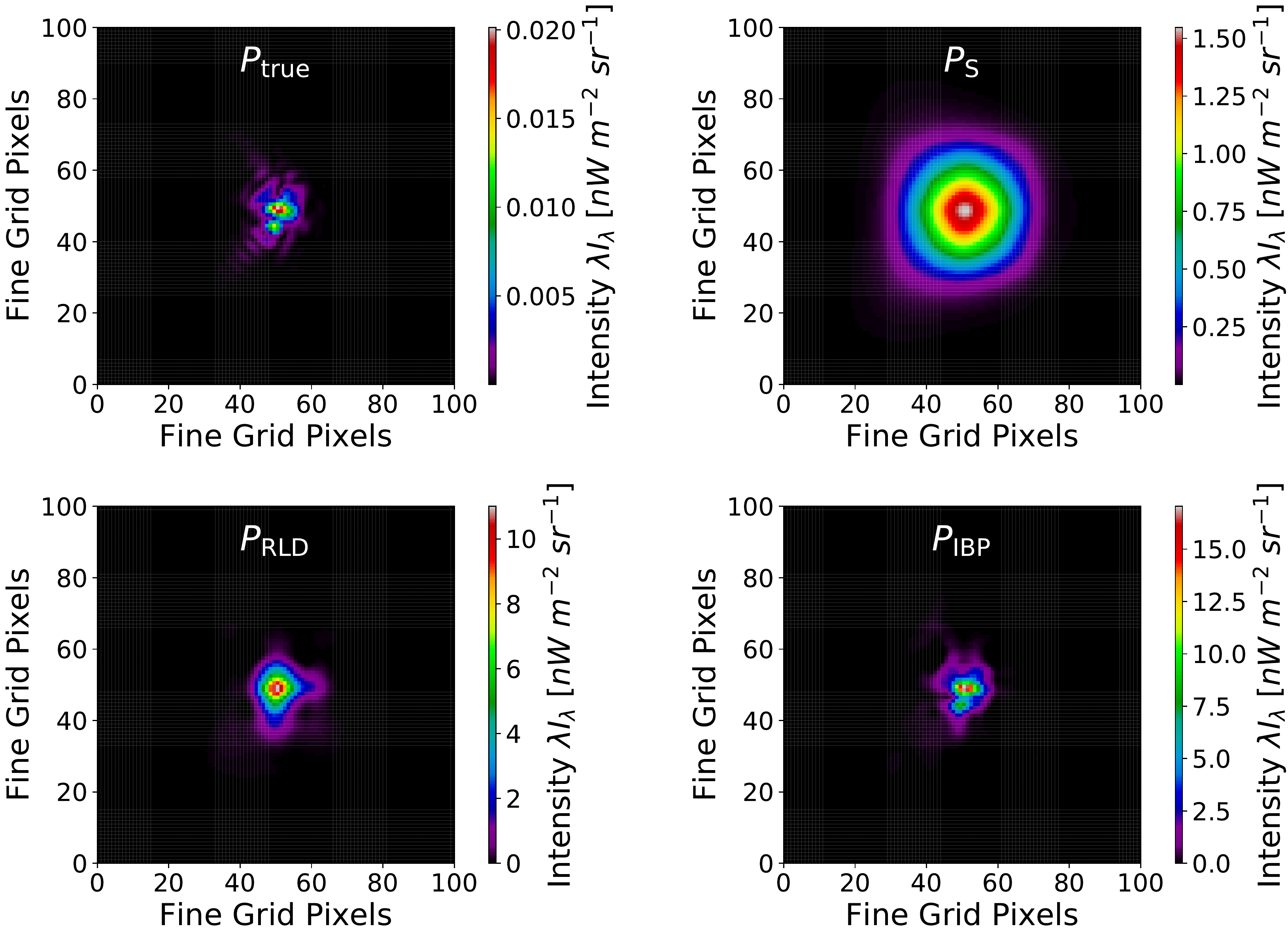}
\caption{The deconvolution process in the presence of noise for the 0.93 $\mu$m SPHEREx wavelength band. In the upper-left panel we show $P_{\rm true}$ for comparison with the noisy $P_{\rm S}$ on the top right, the initial $P_{\rm RLD}$ with the use of damping and background noise constant on the bottom left, and the final $P_{\rm IBP}$ after the IBP process on the bottom right. 
Due to the large number of high-S/N sources in the stack, instrument noise has minimal effect on reconstruction quality. \label{fig:noisy_decon}}
\end{figure*}

Next, we test the reconstruction of the SPHEREx PSF in the presence of noise in the stacked sources. We find that setting pixel values beyond an exclusion radius of 31 fine-grid pixels to zero after both the initial RLD and the IBP sequence yields the best results in this case. This is because noisy pixels with little or no PSF signal can amplify noise artifacts. The damping radius controls the weight given to pixels between the damping and exclusion radii during the RLD, while the exclusion radius provides a hard cutoff for pixels in the $P_{\rm S}$ image that have no significant effect on the PSF. Radii for damping and exclusion are determined through minimization of the average flux deviation derived from using $P_{\rm IBP}$ as a kernel for optimally weighted photometry, $\langle \delta F \rangle$, and its bias from the known input $F$. Because the average deviation from the expected flux is desired to be zero, any significant trend from zero indicates a bias being introduced during reconstruction or photometry. The value and bias of $\langle \delta F \rangle$ are minimized until the number of iterations, $I_{\rm IBP}$, and the exclusion and damping radii both converge. The radii and number of iterations have converged when no further improvements occur. Figure \ref{fig:noisy_decon} demonstrates the deconvolution process for the SPHEREx PSF in the presence of noise. 

\subsubsection{Evaluating Reconstruction Quality}
\label{S:fom}

In order to evaluate the quality of the reconstruction, as well as its effectiveness when used as a weight kernel for optimal photometry, we develop a figure of merit (FOM) based on $\langle \delta F \rangle$. Our FOM is the RMS of $\delta F$:

\begin{equation}
\label{eq:fom1}
{\sigma_{\delta F} = \sqrt{\frac{\sum\limits_{i}^{N}[(\delta F_{i})^{2}]}{N}}}
\end{equation}
where $\delta F_{i}$ is the difference between measured and expected flux for any individual source, and $N$ is the total number of sources. This gives a measure of the total dispersion in $\delta F$ such that $\sigma_{\delta F} < 1$ indicates photometric results that meet the SPHEREx requirement.

\begin{figure}[htbp!]
\centering
\includegraphics[width=3.4in]{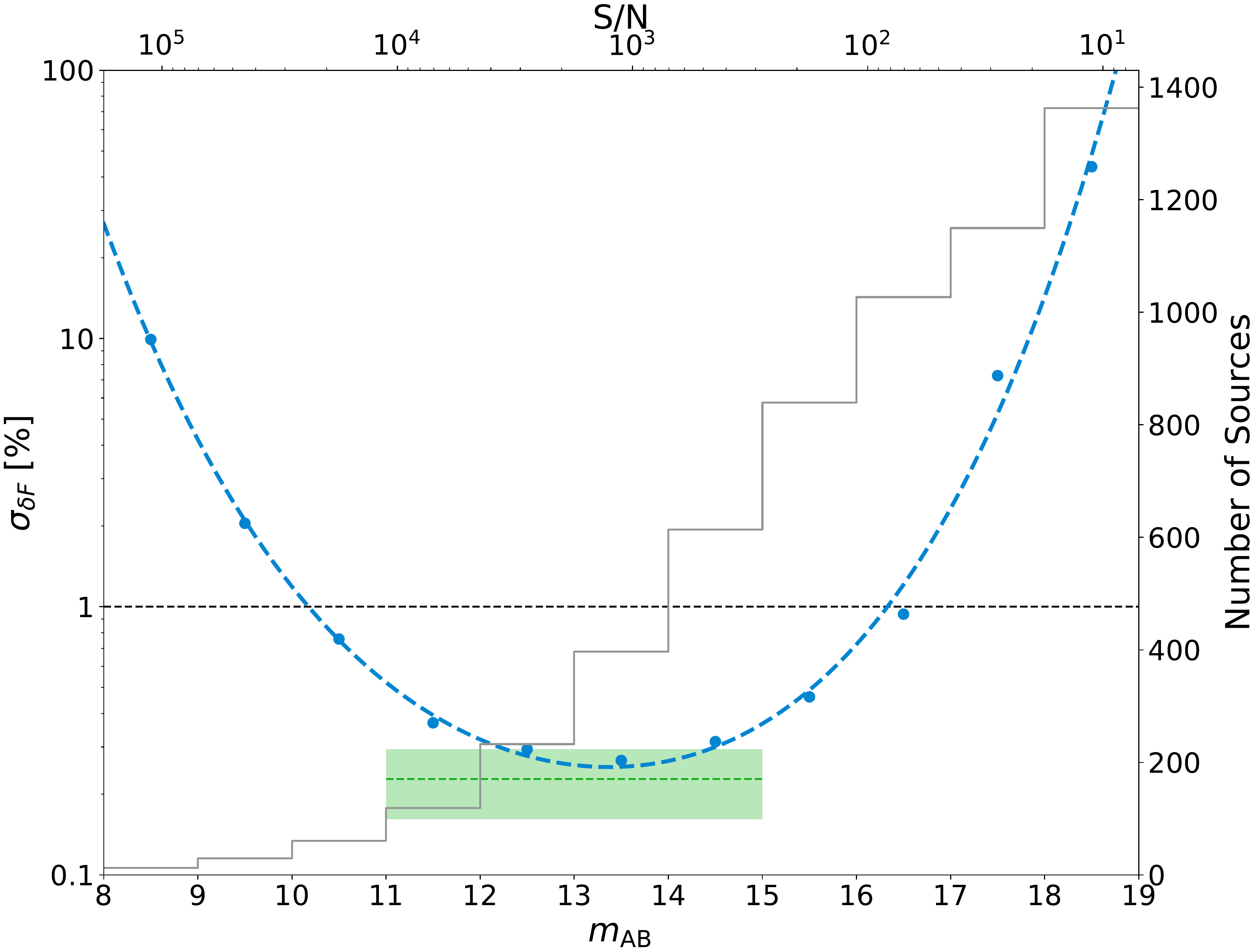}
\caption{Comparison of $\sigma_{\delta F}$ when restricting the sources used in $P_{\rm S}$ to 1 mag wide bins (lower axis) centered between integer $m_{\rm AB}$ values. The blue points and line indicate the average $\sigma_{\delta F}$ (left axis) from 50 trials of each bin at varying galactic longitude and $b$ $>$ 70\degree. The black dashed line shows the desired limit of $\sigma_{\delta F}$ = 1, and the green dashed line indicates the average $\sigma_{\delta F}$ from 50 trials of a $11 \leq m_{AB} \leq 15$ bin, with the green region showing the 1$\sigma$ error. This region provides the lowest $\sigma_{\delta F}$ compared to individual magnitude bins. The gray line shows the number of sources used in $P_{\rm S}$ in any given bin (right axis), and the upper axis shows the corresponding S/N for each magnitude bin. $\sigma_{\delta F}$ always decreases with an increasing number of sources or higher S/N and is minimized in the zone that balances both. \label{fig:mag1}}
\end{figure}

An important issue is the optimal range of sources to use for constructing $P_{\rm S}$. Bright sources allow measurement of $P_{\rm S}$ in the coarse gridding with high fidelity, but as there are relatively few sources, this comes at the cost of the spatial resolution of the reconstruction. Faint sources are numerous, but the effect of noise is larger in the stack, and this has a cascading effect on the fidelity of $P_{\rm IBP}$. In order to optimize the range of fluxes to use, we calculate $P_{\rm IBP}$ over narrow magnitude ranges and from these calculate $\sigma_{\delta F}$ to determine where it is optimal.
Figure \ref{fig:mag1} demonstrates how the average $\sigma_{\delta F}$ of 50 trials with randomized galactic longitude for $b$ $>$ 70\degree \ changes when selecting only sources within a single AB magnitude bin for use in $P_{\rm S}$. We find that $\sigma_{\delta F}$ is minimized for $11 \leq m_{\rm AB} \leq 15$, which are the statistically optimal source magnitudes for optimal PSF reconstruction at the SPHEREx noise amplitude. 

\begin{figure}[htbp!]
\centering
\includegraphics[width=3.3in]{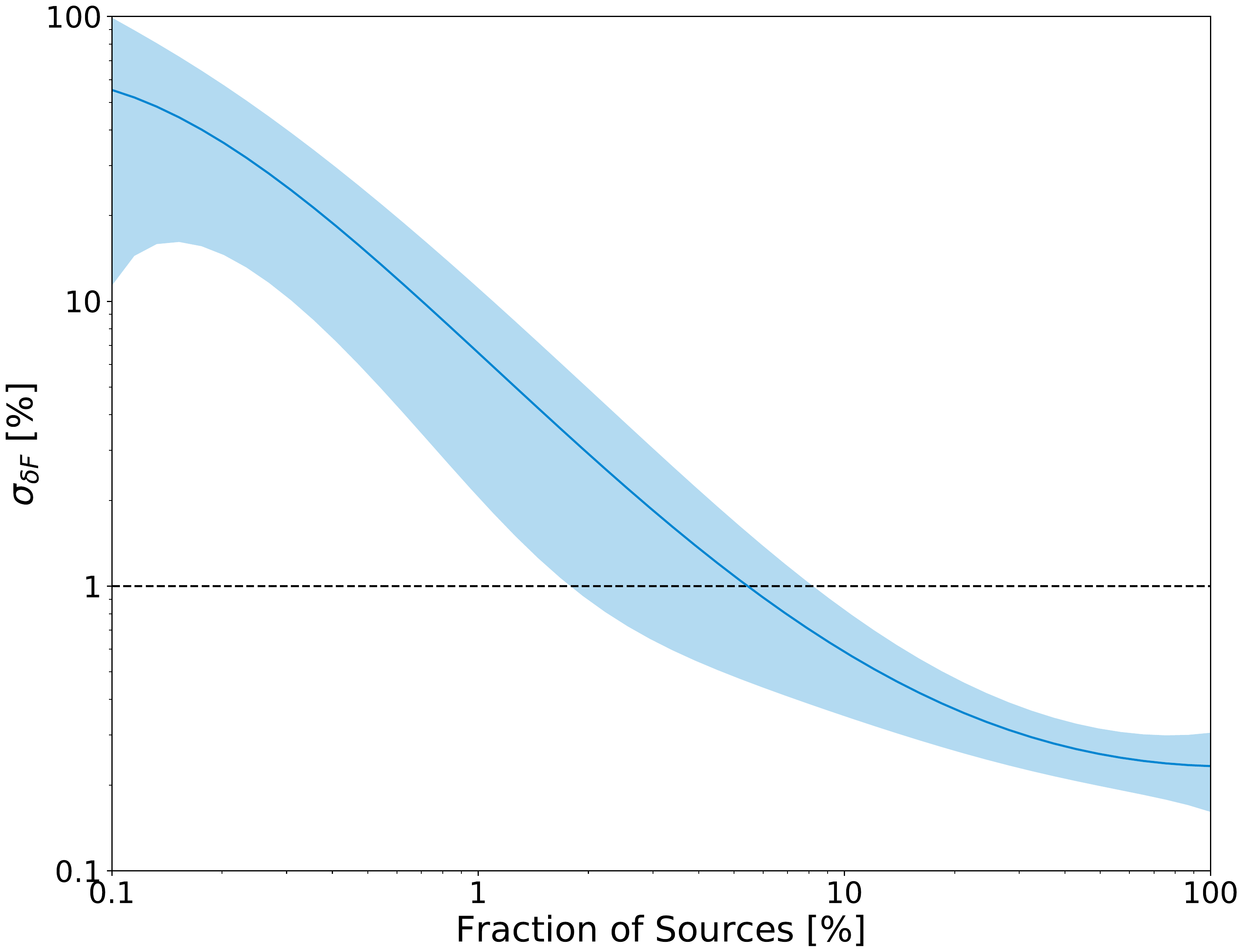}
\caption{Comparison of $\sigma_{\delta F}$ correlating to the fraction of the total number of available sources being used in $P_{\rm S}$. The blue region corresponds to the 1$\sigma$ error from 50 trials of randomized galactic longitude at $b$ $>$ 70\degree. Increasing the number of sources used in $P_{\rm S}$ results in lower $\sigma_{\delta F}$ and thus better reconstruction quality, while using as few as 6\% of sources still allows for $\sigma_{\delta F}$ $<$ 1, marked by the black dashed line. \label{fig:fomperc}}
\end{figure}

 Next, we test how using a much smaller subset of the total available sources affects the reconstruction quality. We restrict the number of sources used in $P_{\rm S}$ by an increasing percentage via a representative sample for the optimal range of $11 \leq m_{AB} \leq 15$. Figure \ref{fig:fomperc} shows how the average $\sigma_{\delta F}$ varies with this restriction for a series of 50 trials of fields with varying galactic longitude and $b > 70\degree$ at each fraction of the total number of sources. As the fraction of sources being used increases, $\sigma_{\delta F}$ continues to decrease, indicating better reconstruction quality. Even restricting to only 6\% of the available sources yields $\sigma_{\delta F} < 1$ on the mean. We conclude that it should be possible to produce accurate kernel reconstructions over regions as small as 10\% of the full SPHEREx FoV. Because the SPHEREx PSF is expected to change somewhat over the FoV as the effective bandpass at the detector changes, this capability is invaluable to produce $P_{\rm S}$ appropriate to a region where the gradient of the PSF properties will be smaller.

\subsubsection{Overall Photometric Accuracy}

\begin{figure*}[htbp!]
\centering
\includegraphics[width=\textwidth,height=0.91\textheight,keepaspectratio]{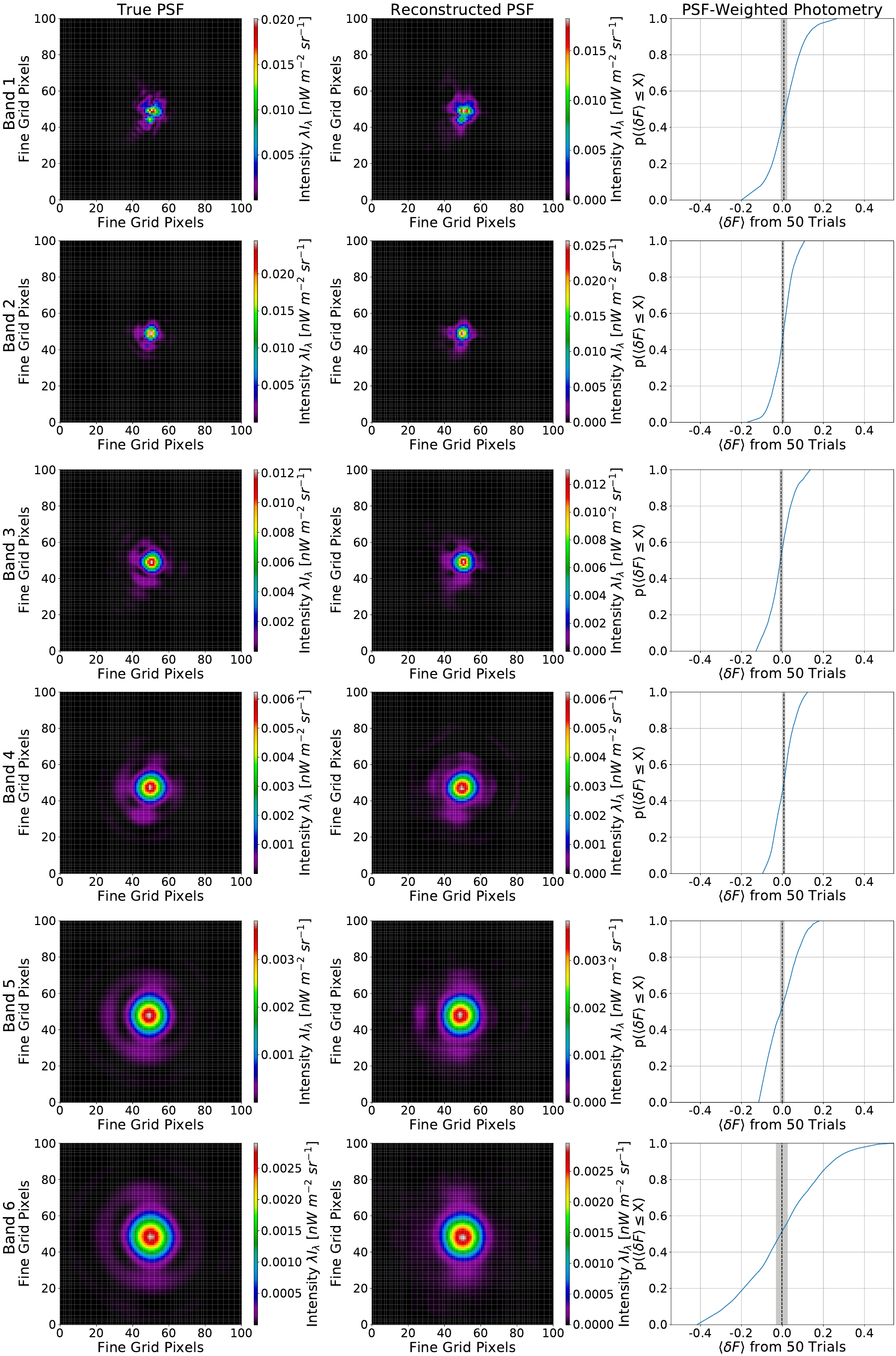}
\caption{Comparison of the six SPHEREx wavelength bands, with each band's $P_{\rm true}$ in the first column, final reconstructed $P_{\rm IBP}$ in the second column, and CDF of $\langle \delta F \rangle$ from 50 trials of randomized galactic longitude at $b$ $>$ 70\degree \ in the third column. Mean $\langle \delta F \rangle$ for all trials is marked with a dotted line, with a gray region showing the standard error of the mean. The mean $\langle \delta F \rangle$ is consistent with zero for each band. \label{fig:sc_noise}}
\end{figure*}

To assess the overall reliability of the IBP algorithm, in Figure \ref{fig:sc_noise} we compare all six SPHEREx wavelength bands (centered at 0.930, 1.375, 2.030, 3.050, 4.065, and 4.730 $\mu$m) using the $11 \leq m_{AB} \leq 15$ range for $P_{\rm S}$. $P_{\rm true}$ and the noisy $P_{\rm IBP}$ are shown for each band, along with a cumulative distribution function (CDF) of the resulting $\langle \delta F \rangle$ for the noisy $P_{\rm IBP}$ applied to noiseless sources for 50 trials with randomized galactic longitude at $b > 70\degree$. We use catalog-based fluxes from the Gaia catalog for all bands, resulting in a high level of accuracy with mean percent error consistent with zero for each band. This demonstrates the algorithm's robustness in successful PSF reconstruction of complex PSFs with different types of structure from images with varying levels of noise. 

\begin{figure*}[htb!]
\centering
\includegraphics[width=7in]{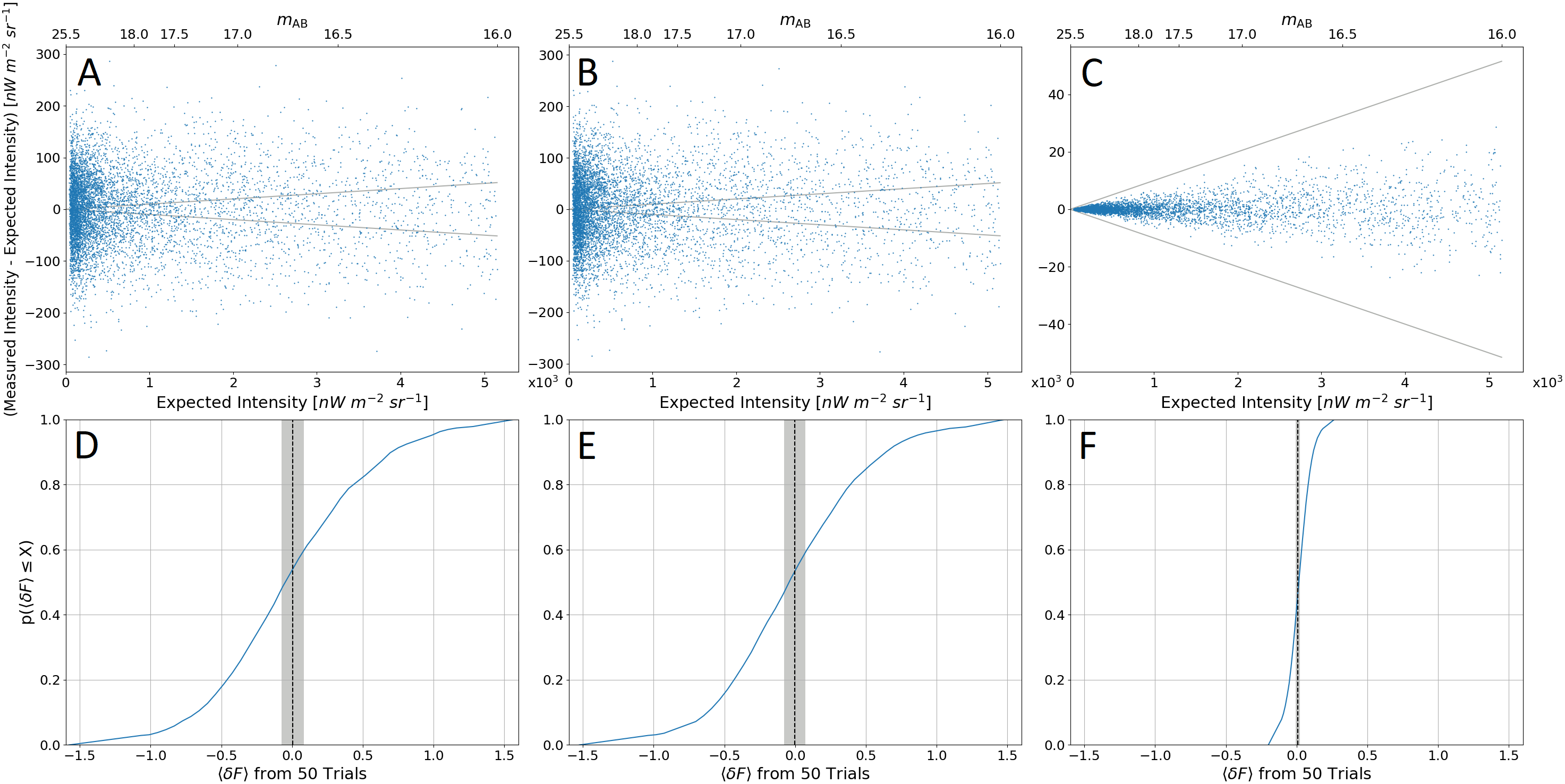}
\caption{The top row shows single trials at the NGP using the SPHEREx 0.93 $\mu$m $P_{\rm true}$ with catalog-based flux values and instrument noise. Panels A and B show photometry resulting from noisy sources with a noisy $P_{\rm IBP}$ and noisy sources with a noiseless $P_{\rm true}$, respectively. In both cases, flux values display similar scatter, indicating that reconstruction quality does not negatively affect photometric results. Panel C shows the results from noiseless sources and a noisy $P_{\rm IBP}$, all of which are within 1\% boundaries. To quantify the bias in these cases, the bottom row shows CDFs of $\langle \delta F \rangle$ corresponding to panels A -- C for 50 trials with randomized field galactic longitude and $b$ $>$ 70\degree, where the dotted line indicates the mean $\langle \delta F \rangle$ of all trials, and the gray region shows the standard error of the mean. Panel D shows a mean $\langle \delta F \rangle$ of $0.005\% \pm 0.080$\%, panel E has a mean $\langle \delta F \rangle$ of $-0.003\% \pm 0.075$\%, and panel F has a mean $\langle \delta F \rangle$ of $0.008\% \pm 0.016$\%.
\label{fig:sphx_noise}}
\end{figure*}

In previous tests, we have performed optimal photometry on intrinsically noiseless sources to assess the accuracy of the IBP reconstruction algorithm. It is also informative to investigate at which source fluxes the instrumental and shot noise present in real sources would dominate over the error arising from misestimation of $P_{\rm true}$. In the top row of Figure \ref{fig:sphx_noise}, we show tests using combinations of noise in the stack and noise in the photometered sources (or lack thereof) for the NGP, compared to using a noiseless $P_{\rm true}$ as the photometry kernel. We continue to restrict $P_{\rm S}$ to $11 \leq m_{AB} \leq 15$. As expected, only when the noisy $P_{\rm IBP}$ is used for photometry on sources with perfectly known flux is the resulting photometry within 1\% variation at all $F$. Photometry on noisy sources is dominated by the noise in the source flux measurement. Noise in the PSF reconstruction itself appears to have a small effect compared with the intrinsic scatter of $F$ from random noise. Additionally, the PSF reconstruction is not correlated with details of the noise realization; the $P_{\rm IBP}$ derived from one noise realization, when used for photometry on sources from different noise realizations, returns an unbiased $\langle \delta F \rangle$.

To understand the relative effect of noise in the stack versus numerical inaccuracies in the algorithm, we perform 50 trials with fields of varying galactic longitude at $b$ $>$ 70\degree \ for the same set of noise cases. We again find that the variation in the output flux is dominated by the intrinsic noise rather than the error in the IBP, demonstrated in the bottom row of Figure \ref{fig:sphx_noise}. Results are very similar for noisy sources with a noiseless $P_{\rm true}$ and noiseless sources with a noisy $P_{\rm IBP}$, as both have mean $\langle \delta F \rangle$ consistent with zero and similar distribution widths. Again, the PSF kernel reconstruction is not introducing photometric error beyond that determined by the noise on the sources.

\subsubsection{Source Crowding}
\label{s:crowd}

To understand the effects of source crowding (discussed in Section \ref{S:stack}), we vary galactic coordinates in a single trial for each of $b$ = 15\degree, 30\degree, 60\degree, and $\ell$ = 0\degree, 90\degree, and 180\degree\ with $P_{\rm S}$ restricted to $11 \leq m_{AB} \leq 15$. All trials use the SPHEREx $P_{\rm true}$ for the 0.93 $\mu$m wavelength band with catalog-based fluxes and noisy $P_{\rm IBP}$ with noiseless sources during photometry. Table \ref{table:galtab} gives a comparison of $\langle \delta F \rangle$ and $\sigma_{\delta F}$ for each set of coordinates. All fields have $\sigma_{\delta F} < 1$ with no bias in $\langle \delta F \rangle$ and no obvious increase in $\sigma_{\delta F}$. This indicates that accurate PSF reconstruction and photometry can be achieved with this method, even for highly crowded fields. 

\begin{deluxetable}{ccccccc}[htbp!]
\tablecaption{Photometric Results as a Function of Sky Position \label{table:galtab}}
\tablehead{\colhead{} & \multicolumn{2}{c}{\boldmath{$\ell = 180\degree$}} & \multicolumn{2}{c}{\boldmath{$\ell = 90\degree$}} & \multicolumn{2}{c}{\boldmath{$\ell = 0\degree$}} \\
\colhead{} & \colhead{\textbf{$\langle \delta F \rangle$}} & \colhead{\textbf{$\sigma_{\delta F}$}} & \colhead{\textbf{$\langle \delta F \rangle$}} & \colhead{\textbf{$\sigma_{\delta F}$}} & \colhead{\textbf{$\langle \delta F \rangle$}} & \colhead{\textbf{$\sigma_{\delta F}$}}}
\startdata
 \boldmath{$b$} \bf{= 60\degree} & 0.031\% & 0.193\% & 0.010\% & 0.163\% & --0.017\% & 0.218\% \\
 \boldmath{$b$} \bf{= 30\degree} & --0.046\% & 0.201\% & 0.004\% & 0.168\% & --0.026\% & 0.207\% \\
 \boldmath{$b$} \bf{= 15\degree} & --0.078\% & 0.205\% & --0.066\% & 0.172\% & --0.090\% & 0.210\% \\
\enddata
\end{deluxetable}

\vspace{30pt}
\subsection{LORRI PSF Estimation}
\label{S:lorri}

\begin{figure*}[htb!]
\centering
\includegraphics[width=6.5in]{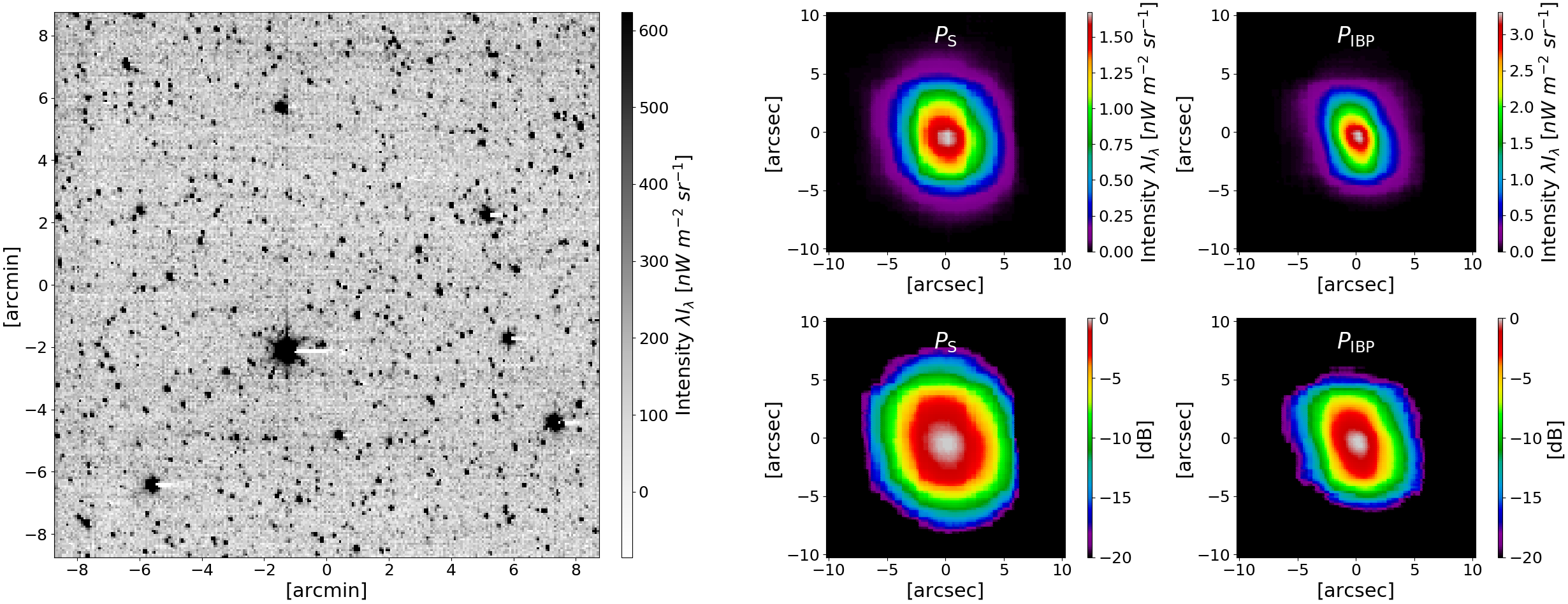}
\caption{On the left is an example LORRI image with 4.$^{\prime \prime}$1 pixel$^{-1}$. On the right is a comparison of the $P_{\rm S}$ and $P_{\rm IBP}$ measured from this image, shown in linear scale in the top row and logarithmic scale in the bottom row. The PSF measured in this image is significantly larger than the laboratory-determined PSF \citep{Cheng2008}, which is likely due to the pointing stability of the spacecraft \citep{Noble2009}. \label{fig:lorri}}
\end{figure*}

To this point, we have been working solely with simulated PSFs where we know the input and the noise is idealized. To test this algorithm against real data, we operate on an image taken by the LORRI instrument on New Horizons \citep{Cheng2008, Zemcov2017}. This image was acquired by the CCD chip on the LORRI instrument and, in LORRI's 4 $\times$ 4 binning mode, has size $256 \times 256$ pixels at 4$.^{\prime \prime}$1 pixel$^{-1}$, as shown in Figure \ref{fig:lorri}. In the PSF reconstruction, we use $r = 20$ and $I_{\rm IBP} = 10$. The reconstructed PSF is shown in Figure \ref{fig:lorri}. We find that the FWHM of the minor axis of $P_{\rm IBP}$ is $3.^{\prime \prime}73$, while the FWHM of the major axis is $5.^{\prime \prime}51$, both of which are significantly larger than the $1.^{\prime \prime}5$ PSF measured in the lab \citep{Cheng2008}. 

What can account for this difference? \citet{Noble2009} demonstrate that the PSF width is a strong function of the integration time of the instrument, and the New Horizons spacecraft is known to exhibit pointing drift at the arcsec s$^{-1}$ level. Performing the deconvolution on a set of several hundred 10 s LORRI exposures, we find the PSF is often extended with eccentricity $\epsilon > 0.8$ and a minimum pointing drift of $1^{\prime \prime}$. This is consistent with the expected drift in the spacecraft's pointing in relative control mode of $\pm 2^{\prime \prime}$ per exposure \citep{Conard2017}. In principle, if we understood the pointing history of the spacecraft, we could deconvolve this component out to isolate the underlying optical PSF using the same methods described above; this will be left to future work.

\section{Discussion}
\label{S:discussion}

\subsection{Comparison to Similar Methods}
\label{sS:similarmethods}

The method of PSF reconstruction described here is robust to complicating factors such as severe undersampling, complex PSFs, noise, crowded fields, or a limited number of available point sources. \citealt{decon_review} and \citealt{dig_im_recon} provide thorough reviews of deconvolution methods commonly used in astronomy, which are numerous. New methods are still being actively developed for applications in astronomy \citep[see, e.g.,][]{decon_ml}. In comparison to some algorithms, a major advantage of this method is that it works on a per-exposure basis. Many other superresolution methods that have been used in astronomy (e.g.,~\citealt{sr_spire,sr_ground,sr_unreg}) rely on multiple exposures of the same field for information reconstruction. As an example, data from the Hubble Space Telescope (HST) prompted the development and application of a number of techniques driven by undersampling in the Wide Field and Planetary Camera 2 (WFPC2; \citealt{Anderson_2000}) and Wide Field Camera 3 (WFC3; \citealt{wfc3_models}). These methods rely on dither-based solutions in which multiple images of the same field with some small shift in the position of the detector are recombined to generate a higher-resolution image of the same field \citep{Lauer1999,Drizzle}. Our method does not rely on similar constraints. 

The SPHEREx application investigated here requires superresolution knowledge of the PSF to optimally weight pixels for photometry and, as a result, requires the effect of the pixelization to be modeled on a per-source basis. 
Some reconstruction methods return the PSF convolved with the pixel-gridding function (the effective PSF or the PSF observed in the image) instead of the underlying optical PSF, such as the method introduced by \cite{Anderson_2000}. While the effective PSF has important uses, it cannot be used directly for optimal photometry, pointing drift assessment, or other applications where the details of the underlying optical PSF are important. Other methods of superresolution image reconstruction can also deconvolve the pixel-gridding function and be applied to single exposures \citep{Aujol2006}, but have applications significantly different from those discussed here (e.g.,~\citealt{sr_euclid}).

Computational speed is also an advantage of this method. More complicated methods such as blind deconvolution \citep{blind_decon} or Bayesian methods \citep{ibp_comp} have also been used for analyses in astronomy, but these tend to be much more computationally expensive. Comparatively, our method of stacking and deconvolution requires $\sim$1 minute to analyze a single exposure on a personal computer, which is scalable to large surveys. 

\subsection{Future Improvements}
\label{sS:performance}

To summarize the primary result of this work, through simulations of the stacking method in all six
 SPHEREx wavelength bands including realistic noise, source catalogs from
 Gaia+AllWISE star catalogs, and realistic beam shapes, we find
 IBP-derived kernels allow photometry with accuracy to $\sim$0.2\% in a single SPHEREx exposure. We 
 find that kernels derived from stars with $11 \leq m_{\rm AB} \leq 15$
 generate the best estimate of the underlying optical PSF across all six
 SPHEREx bands. At SPHEREx's noise level, the population of sources
 in this range balance the need for large S/N in the stack against
 the need for a large number of sources to maintain spatial fidelity.
We find that stacking on subimages of the full SPHEREx FoV does not significantly degrade the performance of the
 algorithm up to fields with $\lesssim$10\% of the full FoV, at least at high and mid-galactic latitudes. 
 
Though we have determined a method that meets the 1\% accuracy requirement, further improvements and details could be investigated. The IBP reconstruction method is primarily tuned to returning a kernel for optimal photometry of point sources, rather than reconstruction over a wide range of spatial scales. ``Stitching'' together information from the brightest sources to probe the faint wings of the PSF with measurements of the central region of the PSF derived using IBP may offer an improved measurement of the PSF over a wide range of spatial scales \citep{Zemcov2014}. 
Further, more advanced algorithms that can separate effects like the optical PSF, pointing jitter and drift, and distortion over the array may be possible using larger volumes of data. Though such advancements might offer even more accurate reconstructions, the method presented here already illustrates how advanced statistical methods can offer a path to unlocking the full information content of astronomical images.

\section*{Acknowledgments} 
Thanks to Chi Nguyen for helpful comments and suggestions.
This work was supported by NASA awards 80GSFC18C0011/S442557, NNN12AA01C/1594971, and 80NSSC18K1557.
The research was partly carried out at the Jet Propulsion Laboratory, California Institute of Technology, under a contract with NASA (80NM0018D0004).

This publication makes use of data products from the Wide-field Infrared Survey Explorer, which is a joint project of the University of California, Los Angeles, and the Jet Propulsion Laboratory/California Institute of Technology, and NEOWISE, which is a project of the Jet Propulsion Laboratory/California Institute of Technology. WISE and NEOWISE are funded by the National Aeronautics and Space Administration.
This work has made use of data from the European Space Agency (ESA) mission Gaia (\url{https://www.cosmos.esa.int/gaia}), processed by the Gaia Data Processing and Analysis Consortium (DPAC, \url{https://www.cosmos.esa.int/web/gaia/dpac/consortium}). Funding for the DPAC has been provided by national institutions, in particular the institutions participating in the Gaia Multilateral Agreement.

The authors acknowledge Research Computing at the Rochester Institute of Technology for providing computational resources and support that have contributed to the research results reported in this publication.

\software{Astropy \citep{2013A&A...558A..33A, 2018AJ....156..123A}, Matplotlib \citep{Hunter:2007}, NumPy \citep{van2011numpy}, SciPy \citep{Virtanen_2020}.}
\newpage
\bibliography{psf_simulation}

\appendix

\section{Point-source Flux and Units}
\label{S:units}

For point sources that are assigned some magnitude and corresponding specific flux $F_{\lambda}$ (measured in units of power per unit area per unit wavelength such as nW m\textsuperscript{-2} $\mu$m\textsuperscript{-1}), we calculate specific intensity (generally measured in units of power per unit area per unit solid angle per unit wavelength) $I_{\lambda}$ as
\begin{equation}
\label{eq:specint}
{I_{\lambda} = \frac{F_{\lambda}}{\Omega_{\rm beam}}\ [\textnormal{nW}\ \textnormal{m}^{-2}\ \textnormal{sr}^{-1}\ \mu \textnormal{m}^{-1}]}
\end{equation}
where $\Omega_{\rm beam}$ is the beam area on the sky for SPHEREx. We then multiply $I_{\lambda}$ by a specific wavelength to get the intensity or diffuse surface brightness $\lambda I_{\lambda}$, where
\begin{equation}
\label{eq:int}
{\lambda I_{\lambda} = \frac{\lambda F_{\lambda}}{\Omega_{\rm beam}}\ [\textnormal{nW}\ \textnormal{m}^{-2}\ \textnormal{sr}^{-1}]}.
\end{equation}
The quantity $\lambda I_{\lambda}$ is equivalent to the quantity $\nu I_{\nu}$, where $I_{\nu}$ is measured in units of power per unit area per unit solid angle per unit frequency such as nW m\textsuperscript{-2} sr\textsuperscript{-1} Hz\textsuperscript{-1}. With the proper conversion, $I_{\lambda}$ and $I_{\nu}$ can be used interchangeably. When we perform photometry of a point source, we are taking the integral
\begin{equation}
\label{eq:phot}
{\int_{\rm beam}I_{\nu}d\Omega = F_{\nu}\ [\textnormal{Jy}]}
\end{equation}
to get some flux $F_{\nu}$. For our point sources with catalog-based flux values, we choose to convert the photometric flux into intensity in units natural to SPHEREx, nW m\textsuperscript{-2} sr\textsuperscript{-1}, using equation \ref{eq:int}. Given some source with intensity $\lambda I_{\lambda}$ and a pixel RMS $\sigma$, the S/N can be defined as
\begin{equation}
\label{eq:S/N}
{{\rm S}/{\rm N} = \frac{\lambda I_{\lambda}}{\sigma}}
\end{equation}
where $\lambda I_{\lambda}$ and $\sigma$ have been integrated over a beam with area on the sky $\Omega_{\rm beam}$. For our simulated point sources with uniform flux, we assign $\lambda I_{\lambda}$ such that the S/N is a desired value based on the selected pixel RMS, and quote the S/N as scaled flux of dimensionless units, easily converted into flux units via equations \ref{eq:S/N} and \ref{eq:int}. 

\end{document}